\documentclass[11pt,a4paper]{article}
\pdfoutput=1
\usepackage{jheppub}
\usepackage[utf8]{inputenc}
\usepackage[british]{babel}
\usepackage{lmodern}
\usepackage{graphicx}
\usepackage{epstopdf}
\usepackage{slashed}
\usepackage{tikz}
\usepackage{tikz-feynman}
\usepackage{float}

\newcommand{\beq}{\begin{equation}}
\newcommand{\eeq}{\end{equation}}
\newcommand{\bea}{\begin{eqnarray}}
\newcommand{\eea}{\end{eqnarray}}
\newcommand{\nn}{\nonumber}

\newcommand{\fig}{figure~}

\newcommand{\bx}{{\bf x}}

\newcommand{\bp}{{\bf p}}

\newcommand*\diff{\mathop{}\!\mathrm{d}}

\def\lsi{\raise0.3ex\hbox{$<$\kern-0.75em\raise-1.1ex\hbox{$\sim$}}}
\def\gsi{\raise0.3ex\hbox{$>$\kern-0.75em\raise-1.1ex\hbox{$\sim$}}}

\graphicspath{{./figures/}}

\preprint{DESY 20-065}

\title{Complete leading-order standard model corrections to quantum leptogenesis}

\author[a]{Paul Frederik Depta,} 
\author[b]{Andreas Halsch,}
\author[b]{Janine H\"utig,} 
\author[c]{Sebastian Mendizabal,}
\author[b]{and Owe Philipsen}

\affiliation[a]{Deutsches Elektronen-Sychrotron DESY, Notkestra\ss e 85, D-22607 Hamburg, Germany}
\affiliation[b]{Institut f\"ur Theoretische Physik,
Goethe-Universit\"at Frankfurt am Main, \\ Max-von-Laue-Str. 1, 60438 Frankfurt am Main, Germany}
\affiliation[c]{Department of Physics, Universidad T\'ecnica Federico Santa Mar\'ia,\\ Casilla 110-V, Valpara\'iso, Chile} 

\abstract{
Thermal leptogenesis, in the framework of the standard model with three additional heavy Majorana neutrinos, provides an attractive
scenario to explain the observed baryon asymmetry in the universe. It is based on the out-of-equilibrium 
decay of Majorana neutrinos in a thermal bath of standard model particles, 
which in a fully quantum field theoretical formalism is obtained by solving 
Kadanoff-Baym equations. So far, the leading two-loop contributions from leptons and Higgs particles 
are included, but not yet gauge corrections. These enter at three-loop level but, 
in certain kinematical regimes, require a resummation to infinite loop
order for a result to leading order in the gauge coupling. 
In this work, we apply such a resummation to the calculation of the lepton number density. 
The full result for the simplest ``vanilla leptogenesis'' scenario 
is by $\mathcal{O}(1)$ increased compared to that of quantum Boltzmann equations, and for 
the first time permits an estimate of all theoretical uncertainties.
This step completes the quantum theory of leptogenesis and forms the basis for quantitative evaluations, as well
 as extensions to other scenarios.
 }

\emailAdd{frederik.depta@desy.de}
\emailAdd{halsch@itp.uni-frankfurt.de}
\emailAdd{janine.schaefer1@web.de}
\emailAdd{sebastian.mendizabal@usm.cl}
\emailAdd{philipsen@itp.uni-frankfurt.de}

\begin{document}
\maketitle


\section{Introduction}
The origin of the observed baryon asymmetry in the universe is an as yet 
unsolved problem of physics. Although  
the standard model of particle physics (SM) features baryon plus lepton number ($B+L$) 
and $CP$-violating processes before the electroweak phase transition~\cite{Kuzmin:1985mm}, 
the amount of $CP$ violation is too small to arrive at the observed 
asymmetry \cite{Huet:1994jb}. Moreover, for the experimentally observed Higgs 
mass of $\sim$ 125 GeV \cite{Patrignani:2016xqp}, the electroweak phase transition is 
merely an analytic crossover \cite{Buchmuller:1994qy,Kajantie:1995kf,Csikor:1998ew}, 
whereas a sufficiently strong first-order transition is required to provide 
a departure from equilibrium for baryogenesis.
Viable models for baryogenesis are thus based on extensions of the SM, 
with, e.g., additional 
particles on a GUT scale, which then generate a finite asymmetry via $CP$-violating 
out-of-equilibrium decays.

A particularly attractive model in this context is thermal 
leptogenesis \cite{Fukugita:1986hr}. In this case, the SM is extended 
by three additional right-handed Majorana neutrinos, 
with Yukawa couplings to the SM Higgs field and left-handed leptons. 
In the standard scenario~\cite{Blanchet:2012bk}, these Majorana neutrinos are 
produced in the early universe at temperatures beyond their mass scale, $T>M$. When the 
temperature drops below their mass, they decay out-of-equilibrium violating
$CP$, such that a lepton asymmetry is generated. Additional washout processes 
diminish this asymmetry, until they fall out of equilibrium and a finite 
lepton asymmetry is frozen in. Any lepton asymmetry is partially converted into a baryon 
asymmetry via SM sphaleron transitions, which violate $B+L$ while 
keeping $B-L$ constant \cite{tHooft:1976rip,tHooft:1976snw}.
Besides giving robust predictions for baryogenesis, this simple
extension of the SM would also explain the smallness of the light neutrino masses 
via the see-saw mechanism \cite{PhysRevLett.44.912}. For recent reviews, also including other 
scenarios for leptogenesis, see \cite{Biondini:2017rpb,Drewes:2017zyw,Chun:2017spz,Hagedorn:2017wjy,Dev:2017trv,Dev:2017wwc}.

Since leptogenesis is a non-equilibrium process, it is frequently studied using 
Boltzmann equations, which are inherently classical.
For the collision terms, zero temperature 
matrix elements are used, supplied with thermal momentum distribution 
functions which introduce quantum effects. 
However, also off-shell and memory effects become important for 
non-equilibrium processes and have to be treated appropriately.
Further conceptual difficulties
arise in gauge theories, where gauge boson numbers are neither gauge invariant
nor conserved. 
The problem of including quantum effects by hand into a classical framework
is avoided by a formal, quantum field theoretical treatment based on Kadanoff-Baym 
equations, which contain everything once computed to sufficient depth in perturbation theory. 
In \cite{Anisimov:2010dk},
quantum corrections to the Majorana neutrino reaction rates due to Higgs and lepton loops have been included,
and a detailed comparison with the solutions of Boltzmann equations was given.   
Qualitative agreement was achieved by additionally introducing thermal damping widths for Higgs and lepton propagators,
which are expected to be caused by interactions with the weak gauge bosons and may significantly modify 
the result at a quantitative level \cite{Anisimov:2010aq}.

In this work we complete the quantum mechanical treatment of leptogenesis by calculating
gauge corrections systematically. First entering at three-loop level,
their evaluation is complicated by the dynamical generation of 
scales, which necessitate infinite resummations in order to complete even
a leading-order result in the gauge coupling. Such resummations have been performed for
the Majorana neutrino self-energy and the associated production rate  \cite{Anisimov:2010gy}.
Here we extend these calculations to the lepton number density in a first complete 
quantum mechanical calculation of leptogenesis.
Our formulation using Kadanoff-Baym equations \cite{Anisimov:2010dk}
differs from other field-theoretical approaches, in which rate equations are obtained by
exploiting hierarchies of scales in the non-relativistic regime, $T<M$ \cite{Bodeker:2013qaa,Bodeker:2014hqa,Bodeker:2017deo,Biondini:2016arl}.

In order to render this paper self-contained, 
sections \ref{sec:leptogenesis} and \ref{sec:CTLResummation} 
recall the formalism and previous results, which  
form the basis for our calculation.
In section \ref{sec:cylinder}, we devise a resummation technique to 
systematically include gauge corrections to the lepton number density,
which then is complete to leading order in all SM corrections. 
In section \ref{sec:num}, some approximations are motivated which allow for a 
simple evaluation. 
Finally, we compare to previous results in the literature and discuss the significance of 
gauge corrections.           

\section{Thermal leptogenesis in non-equilibrium quantum field theory}
\label{sec:leptogenesis}

\subsection{The leptogenesis scenario}
\label{sec:scenario}

Starting from the SM, a standard approach to leptogenesis is to add three additional 
right-handed Majorana neutrinos (which are electroweak singlets) to the Lagrangian $\mathcal{L}_\text{SM}$,
\begin{align}
\mathcal{L} = \mathcal{L}_{\text{SM}} + \bar{\nu}_{Ri} \text{i} \slashed \partial \nu_{Ri} + \bar{l}_{Li} \tilde{\phi} \lambda_{ij}^* \nu_{Rj} + \bar{\nu}_{Rj} \lambda_{ij} l_{L i} \phi - \frac{1}{2} M_{ij} \left( \bar{\nu}^{c}_{Ri} \nu_{Rj} + \bar{\nu}_{Rj} \nu^{c}_{Ri} \right)\;.
\end{align}
Here the flavour indices $i,j=1,2,3$ count the families in the gauge eigenbasis, and we 
denote $\nu_{R}^{c} = C \bar{\nu}^T_R$, with $C={\text i}\gamma^2\gamma^0$ 
the charge conjugation matrix, and $\tilde{\phi} = \text{i} \sigma_2 \phi^*$. 
The additional neutrinos are weakly coupled to 
SM Higgs fields $\phi$ and massless left-handed leptons $l_{Li}$ via Yukawa couplings 
$\lambda_{ij}$. 
Without loss of generality, we consider here the simplest scenario, also termed 
``vanilla leptogenesis'', in which 
the final asymmetry is assumed to be independent of its flavour composition and,
in a diagonal mass basis, the Majorana masses $M_i$ are hierarchically ordered, 
$M_{i > 1} \gg M_1 := M$. For a review including details, viable parameter ranges and other scenarios, 
see \cite{Blanchet:2012bk}. 
In this case, the two heavier neutrinos can be integrated out, leading to the effective
Lagrangian for the lightest neutrino $N=N_1=\nu_{R1} + \nu_{R1}^c$,
\begin{align}
\label{eq:effLagrangian}
\mathcal{L} = \mathcal{L}_{\text{SM}} + \frac{1}{2} \bar{N} \text{i} \slashed \partial N +& \bar{l}_{Li} \tilde{\phi} \lambda_{i1}^* N + N^T \lambda_{i1} C l_{Li} \phi - \frac{1}{2} M N^T C N \notag \\ +& \frac{1}{2} \eta_{ij} l_{Li}^T \phi C l_{Lj} \phi + \frac{1}{2} \eta_{ij}^* \bar{l}_{Li} \tilde{\phi} C \bar{l}_{Lj}^T \tilde{\phi}
\end{align}
with an effective vertex for lepton and Higgs fields and a combination of Majorana neutrino couplings, respectively,
\begin{align}
\eta_{ij} = \sum_{k > 1} \lambda_{ik} \frac{1}{M_k} \lambda_{kj}^T\;,~~~~\lambda^2 := \sum_i |\lambda_{i1}|^2\;.
\end{align}
The heavy neutrino mass is in the range
$10^9 \, \mathrm{GeV} \lesssim M \lesssim 10^{15} \, \mathrm{GeV}$ \cite{Blanchet:2012bk}, 
and its coupling is assumed very weak compared to SM couplings $g_{\rm SM}$, $\lambda \ll g_{\rm SM}$. Explicitly, the relevant SM couplings for our work are the $SU(2)$ and $U(1)$ gauge group couplings $g$ and $g'$, the Yukawa coupling of the top quark $h_t$, 
and the Higgs self coupling $\lambda_{\phi}$, i.e.\ $g_\mathrm{SM} \in \{ g, g', h_t, \sqrt{\lambda_\phi} \}$.

During the generation of the lepton asymmetry, the heat bath is kept in thermal equilibrium via SM interactions. 
These act on a time scale $\tau_{\text{SM}} \sim 1/(g_{\rm SM}^2T)$, which is much shorter than the 
equilibration time of the heavy neutrino, $\tau_N \sim 1/(\lambda^2 M)$ for $T \gtrsim M$,
which thus is out of equilibrium.
During this first stage, heavy Majorana neutrinos are produced from zero initial abundance by scattering of leptons and Higgs bosons.
This out-of-equilibrium process generates an initial lepton asymmetry, which is later diminished by washout processes.
At temperatures 
$T \sim M$, part of the Majorana neutrinos decay out of equilibrium to produce a final lepton asymmetry, or $B-L > 0$, which is
of the same size as the initial one \cite{Buchmuller:2002rq}.\footnote{A final lepton asymmetry of similar size is also generated 
when starting from a non-zero Majorana neutrino abundance \cite{Buchmuller:2002rq}.} 
Since one can neglect the Hubble expansion, i.e.~assume a constant temperature, during the first stage, 
we will for technical convenience follow \cite{Anisimov:2010dk} in calculating 
this initial lepton asymmetry at fixed $T$ starting from zero initial Majorana neutrino abundance. 
In reality, any lepton asymmetry gets continually converted to a baryon asymmetry by 
sphaleron processes, which are in equilibrium for temperatures above the electroweak phase transition. 
Our first calculation of gauge corrections focuses on the lepton aymmetry only and neglects these and other spectator processes, 
which may affect the final baryon asymmetry by $\sim$ 50\% \cite{Buchmuller:2001sr,Nardi:2005hs,Garbrecht:2014kda}.

\subsection{Correlation functions and Kadanoff-Baym equations}
\label{sec:CorrFcts}

\begin{figure}[t]
\includegraphics[width=0.5\textwidth]{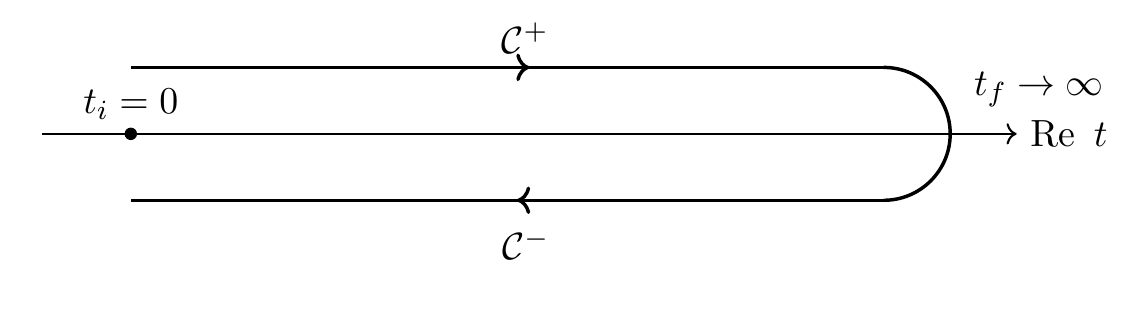}
\includegraphics[width=0.5\textwidth]{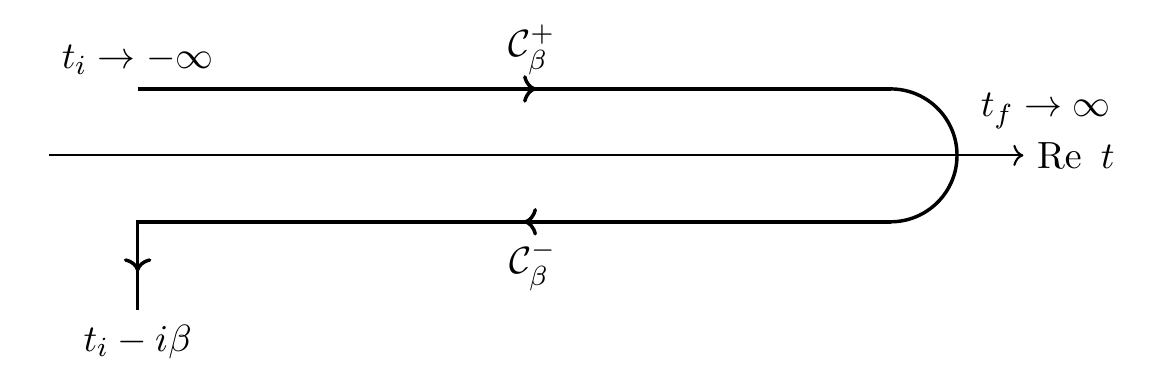}
\caption{Non-equilibrium (left) and equilibrium (right) contours $\mathcal{C}$ in the complex time plane. 
\label{fig:contour}}
\end{figure}
In a quantum field theoretical description of leptogenesis, the desired information
is encoded in the non-equilibrium 
correlation function of the Majorana neutrino, which is weakly coupled to a thermal bath of SM particles.
This requires the real time formalism of thermal quantum field theory \cite{Bellac:2011kqa},
in which two-point Green functions live on a contour $\mathcal{C}$ in the complex 
time plane, whose form depends on whether the field is in equilibrium or not,
cf.~figure~\ref{fig:contour}. 
Our formulation and notation follows \citep{Anisimov:2010dk}, where
details of the calculation can be found.
Green functions can be expressed in terms of Wightman
functions, which for a real scalar field read
\bea
\Delta_{\mathcal{C}}(x_1, x_2) &=& \theta_{\mathcal{C}}(x_1^0, x_2^0)\Delta^>(x_1, x_2) + 
\theta_{\mathcal{C}}(x_2^0, x_1^0)\Delta^<(x_1, x_2)\;,\\
\Delta^>(x_1, x_2) &=& \langle \phi(x_1) \phi(x_2) \rangle\;,\\
\Delta^<(x_1, x_2) &=& \langle \phi(x_2) \phi(x_1) \rangle\;,
\eea
with the Heaviside $\theta_{\mathcal{C}}$-function enforcing path ordering along the time contour $\mathcal{C}$. These in turn are related to the spectral function and statistical propagator, respectively,
\bea
\Delta^-(x_1,x_2)&=&\text{i}\langle [\phi(x_1),\phi(x_2)]\rangle 
=\text{i}\Big(\Delta^>(x_1,x_2)-\Delta^<(x_1, x_2)\Big)\;,\\
\Delta^+(x_1,x_2)&=&\frac{1}{2}\langle \{\phi(x_1),\phi(x_2)\}\rangle 
=\frac{1}{2}\Big(\Delta^>(x_1,x_2)+\Delta^<(x_1, x_2)\Big)\;.
\eea
Analogous quantities are defined for the Green functions of leptons and the 
heavy Majorana neutrino, with the role of (anti-)commutators exchanged for fermions, 
\bea
S^-_{\alpha\beta}(x_1, x_2) &= \text{i} \langle \{ l_{L,\alpha}(x_1),\bar{l}_{L,\beta}(x_2) \} \rangle\;,\quad 
S^+_{\alpha\beta}(x_1, x_2) &= \frac{1}{2} \langle [ l_{L,\alpha}(x_1),\bar{l}_{L,\beta}(x_2) ] \rangle\;,\\
G^-_{\alpha\beta}(x_1, x_2) &= \text{i} \langle \{ N_{\alpha}(x_1),\bar{N}_{\beta}(x_2) \} \rangle\;,\quad 
G^+_{\alpha\beta}(x_1, x_2) &= \frac{1}{2} \langle [ N_{\alpha}(x_1),\bar{N}_{\beta}(x_2) ] \rangle\;.
\eea
In the following, $\Pi_{\mathcal{C}}$ denotes the self-energy of the leptons and 
$\Sigma_{\mathcal{C}}$ that of the Majorana neutrino. These are equally defined on 
the time contour and can be similarly 
decomposed into Wightman functions, 
\bea
\Sigma_{\cal C}(x_1,x_2) = \theta_{\cal C}(x_1^0,x_2^0) \Sigma^>(x_1,x_2) +\theta_{\cal C}(x_2^0,x_1^0) \Sigma^<(x_1,x_2) \;.
\eea

\begin{figure}[t]
\begin{center}
\includegraphics[scale=1.2]{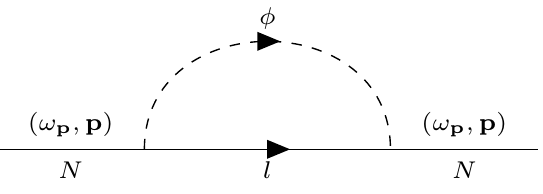}
\caption{One-loop self-energy contribution to the Majorana neutrino self-energy $C \Sigma^{\pm}_{\bp}$. \label{fig:MajoranaSelf}}
\end{center}
\end{figure}

According to the physics scenario given in section \ref{sec:scenario},
the non-equilibrium Green function of the Majorana neutrino has to be evaluated on 
the Schwinger-Keldysh contour, figure~\ref{fig:contour} (left), while the lepton and Higgs
fields are in the thermal bath of SM particles described by the equilibrium contour, 
\fig\ref{fig:contour} (right), with inverse temperature $\beta=1/T$.
The full time evolution of the Majorana neutrino correlation function is given by 
the coupled set of Kadanoff-Baym equations,
\begin{align}
\label{eq:Kadanoff1}
C(\text{i} \gamma^0 \partial_{t_1} - \bp \boldsymbol{\gamma} - M)G^-_{\bp}(t_1, t_2) 
+ \int\limits_{t_1}^{t_2} \diff t' &C \Sigma_{\bp}^-(t_1, t') G_{\bp}^-(t', t_2) = 0\;, \\
\label{eq:Kadanoff2}
C(\text{i} \gamma^0 \partial_{t_1} - \bp \boldsymbol{\gamma} - M)G^+_{\bp}(t_1, t_2) 
- \int\limits_{t_i}^{t_1} \diff t' &C \Sigma_{\bp}^-(t_1, t') G_{\bp}^+(t', t_2) \\ 
	\notag &=  - \int\limits_{t_i}^{t_2} \diff t' C \Sigma_{\bp}^+(t_1, t') G_{\bp}^-(t', t_2)\;,
\end{align}
with the spatial Fourier transforms assuming spatial homogeneity $(\bx = \bx_1 - \bx_2)$,
\bea
G_\bp^\pm(t_1,t_2)&=& \int \diff^3x\; \text{e}^{-\text{i}\bp\bx}G^\pm(t_1,t_2,\bx)\;, \\
\Sigma_\bp^\pm(t_1,t_2)&=& \int \diff^3x\; \text{e}^{-\text{i}\bp\bx}\Sigma^\pm(t_1,t_2,\bx)\;.
\eea
These equations are exact and in particular contain all quantum effects. Interactions
with the plasma are automatically included via the self-energies $\Sigma^{\pm}(x_1,x_2)$, as can be seen via generalised cutting rules.
The leading-order contribution to the 
self-energy is given by the diagram in figure~\ref{fig:MajoranaSelf}.

An analytic solution can be obtained if the Yukawa coupling of the Majorana neutrino 
to the thermal bath is weak, so that back-reactions, being of higher order and
additionally suppressed by the large number of degrees in freedom of the bath, 
can be neglected, and a narrow-width approximation $\Gamma\sim \lambda^2 T \ll T$ 
is justified \cite{Anisimov:2010dk}.\footnote{Note that for $T<M$ one has $\Gamma\sim \lambda^2 M \ll M$.} 
Since the particles in the loop are thermal, the heavy neutrino spectral function is
time-translation invariant,
\beq
G_{\bp}^-(t_1, t_2) = G_{\bp}^-(t_1-t_2)\;.
\eeq
Using a zero abundance of the Majorana neutrinos at initial time $t_i = 0$ and denoting $t_{12}\equiv(t_1+t_2), \Delta t_{12}\equiv (t_1-t_2)$, the solution for the 
Kadanoff-Baym equation is 
\bea
\!\!\!\!\!\!\! G_{\bp}^-(\Delta t_{12}) &=& \left( \text{i} \gamma_0 \cos(\omega_{\bp}
\Delta t_{12}) 
+ \frac{M - \bp \boldsymbol{\gamma}}{\omega_{\bp}} \sin(\omega_{\bp}
\Delta t_{12}) \right) \text{e}^{-\Gamma_{\bp}(\omega_{\bp})|\Delta t_{12}|/2}C^{-1}\;,\\
\!\!\!\!\!\!\! G_{\bp}^+(t_{12}, \Delta t_{12}) &=& -\left( \text{i} \gamma_0 \sin(\omega_{\bp}\Delta t_{12}) - \frac{M - \bp \boldsymbol{\gamma}}{\omega_{\bp}}\cos(\omega_{\bp}\Delta t_{12}) \right) \\ 
\notag& &\times \left[ \frac{1}{2} \tanh\left( \frac{\beta \omega_{\bp}}{2} \right) 
\text{e}^{-\Gamma_{\bp}(\omega_{\bp})|\Delta t_{12}|/2} 
+ f_F(\omega_{\bp})
\text{e}^{-\Gamma_{\bp}(\omega_{\bp})t_{12}/2}\right] C^{-1}\;,
\eea
with $\omega_{\bp} = \sqrt{\bp^2 + M^2}$ and the 
Fermi-Dirac distribution function $f_F(\omega_{\bp})$. 

In general, $\Gamma_\bp$ is the Majorana neutrino decay width resulting from 
its self-energy via
\beq
\label{eq:ConnDecaySelf}
\omega \Gamma_{\bp} (\omega) = 
\frac{1}{4}\text{tr}\left( -\text{i} \slashed p \Sigma_{\bp}^-(\omega) \right)\;.
\eeq
At the level of the calculation in  \cite{Anisimov:2010dk}, this decay width does not include gauge corrections, and for its later 
use in the lepton number matrix  is modeled by a momentum-independent 
constant. In section \ref{sec:gauge} 
we extend the calculation to include gauge corrections, and keep the full momentum-dependence in the lepton number matrix.
Due to its appearance in the Majorana neutrino propagators, we will often need the on-shell value $\Gamma_{\bp}(\omega_{\bp})$, which satisfies
\begin{align}
\label{eq:gammaSym}
\Gamma_{\bp}(\omega_{\bp}) = \Gamma_{\bp}(-\omega_{\bp}) = \Gamma_{-\bp}(\omega_{\bp}).
\end{align}

\subsection{The lepton number matrix}

\begin{figure}[t]
	\begin{center}
	\includegraphics[scale=1.]{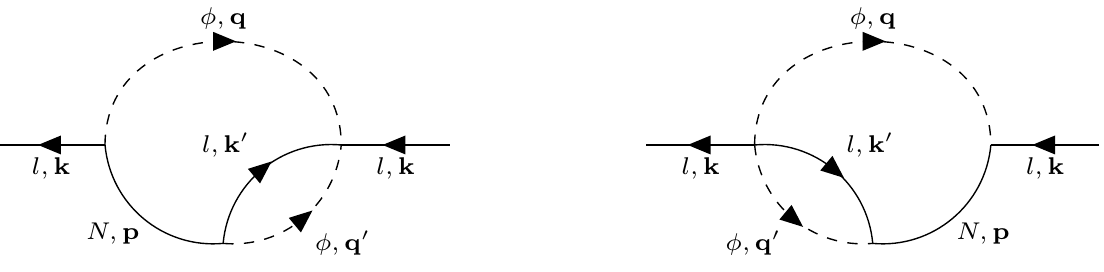}
	\caption{Two-loop corrections to the lepton self-energy $\Pi^{\pm}$ leading to non-zero lepton number densities. \label{fig:CpViolation}}
\end{center}
\end{figure}

A net lepton number density is obtained from the flavour-diagonal  
lepton number current
\bea
n_L&=&\sum_i n_{L,ii}=\frac{1}{V}\int \diff^3x\,\sum_i j^0_{ii}(x)\;,\\
j^\mu_{ij}(x)&=&\bar{l}_{Li}\gamma^\mu l_{Lj}=-\lim_{x'\rightarrow x}\mathrm{tr}\left[ \gamma^{\mu} S_{L,ij}^+(x,x') \right]\;.
\eea
The lepton number matrix is defined as the (spatial) Fourier transform of the 
current's zero component, 
\beq
\label{eq:LeptonNumberDef}
L_{\mathbf{k},ij}(t) = -\text{tr} \left[ \gamma^0 S_{L, \mathbf{k},ij}^+(t,t') \right]_{t \to t'}\;.
\eeq
It has to be evaluated calculating the leading order correction to the lepton statistical propagator, contributing the necessary $CP$-violation. The corresponding Feynman diagrams are shown in figure~\ref{fig:CpViolation}, giving \cite{Anisimov:2010dk}
\begin{align}
\label{eq:nL}
	L_{\mathbf{k},ii} = 12 \, \text{Im}(\lambda_{i1}^*(\eta \lambda^*)_{i1}) \int\limits_0^t \diff t_1 \int\limits_0^t \diff t_2\; \text{Re}\left[ \text{tr} \left( \Pi_{\mathbf{k}}^{(1),>}(t_1, t_2)S^{<}_{\mathbf{k}}(t_2-t_1) \right) \right]\;,
\end{align}
with the lepton self-energy from the first graph\footnote{The contribution of the second 
graph can be expressed in terms of the first one.} in figure~\ref{fig:CpViolation},
\begin{align}
\Pi_{\mathbf{k}}^{(1),>}&(t_1, t_2) = \int\limits_0^{\infty} \diff t_3 \int_{\bp, \mathbf{q}, \mathbf{k}', \mathbf{q}'} (2 \pi)^3 \delta^3(\bp - \mathbf{k}' - \mathbf{q}') (2 \pi)^3 \delta^3(\bp + \mathbf{k} + \mathbf{q}) \\ \notag  
	&\times \left[\tilde{G}_{\bp}(t_1,t_3) \theta(\Delta t_{23}) 
	\left( S_{\mathbf{k}'}^>(\Delta t_{23})\Delta_{\mathbf{q}'}^>(\Delta t_{23}) - S^<_{\mathbf{k}'}(\Delta t_{23})\Delta_{\mathbf{q}'}^<(\Delta t_{23}) \right) 
	\Delta_{\mathbf{q}}^<(\Delta t_{21}) P_L\right]\;.
\end{align}
Here we introduced the notation $\int_\bp\equiv \int \diff^3p/(2\pi)^3$ 
and $\tilde{G}_\bp$ is the scalar non-equilibrium part of the Majorana neutrino propagator, connected to the full propagator via projections
\begin{align}
P_L G_{\bp}(t_1, t_2) C P_L = P_L G_{\bp}^\mathrm{eq}(t_1-t_2) C P_L + \tilde{G}_{\bp}(t_1, t_2) P_L \;,
\end{align}
with the solution
\beq
\tilde{G}_{\bp}(t_1, t_2) = \frac{M}{\omega_{\bp}}\cos(\omega_{\bp} \Delta t_{12}) f_{F}(\omega_{\bp})\text{e}^{-\Gamma_{\bp}t_{12}/2}\;.
\label{eq:MajProp}
\eeq
Note that the equilibrium part of the neutrino propagator does not contribute.  We do not need and hence do not give the explicit form of
 the Higgs and lepton propagators $\Delta_\mathbf{q}^\lessgtr$ and $S_\mathbf{k}^\lessgtr$, which can however be found in~\cite{Anisimov:2010dk}. Using the relations between the different kinds of propagators \citep{Bellac:2011kqa} and identifying
\beq
\Sigma_{\mathbf{k}', \mathbf{q}'}^\lessgtr(\Delta t_{23}) = S^\lessgtr_{\mathbf{k}'}(\Delta t_{23})\Delta_{\mathbf{q}'}^\lessgtr(\Delta t_{23})\;,\quad
\Sigma_{\mathbf{k}, \mathbf{q}}^<(\omega) = -2f_F(\omega)\text{Im} (\Sigma^\mathrm{ret}_{\mathbf{k}, \mathbf{q}}(\omega))\;,
\eeq
it is possible to express the lepton number matrix in terms of the retarded Majorana neutrino self-energy $\Sigma^\mathrm{ret}$. Using 
$\text{Im}(\lambda_{i1}^*(\eta \lambda^*)_{i1}) = 16 \pi \epsilon_{ii}/(3M)$, with $\epsilon_{ii}$ parametrising  
the strength of CP violation \cite{Covi:1996wh}, we have
\bea
\label{eq:TreeLeptonNumberMatrix}
&&L_{\mathbf{k},ii}(t) \\
&=& -\frac{256 \pi \epsilon_{ii}}{M} \int\limits_0^t \diff t_1 \int\limits_0^t  \diff t_2 \int\limits_0^{t_2} \diff t_3 \int\limits_{-\infty}^{\infty}
 \frac{\diff \omega_{21}}{2 \pi} \int\limits_{-\infty}^{\infty} \frac{\diff \omega_{23}}{2 \pi}\int_{\bp, \mathbf{k}', \mathbf{q}, \mathbf{q}'} 
(2 \pi)^6    \delta^3(\bp -\mathbf{k}' - \mathbf{q}') \delta^3(\bp + \mathbf{k} + \mathbf{q} )\nn\\
&&\times \tilde{G}_{\bp} (t_1, t_3)f_F(\omega_{21})
\text{tr} \left( \text{Im} \Sigma_{R, \mathbf{k}, \mathbf{q}}^\mathrm{ret} (\omega_{21}) \text{Im} \Sigma_{L, \mathbf{k}', \mathbf{q}'}^\mathrm{ret} (\omega_{23}) \right) \text{Re} \left[ \text{e}^{-\text{i}(\omega_{21} \Delta t_{21} + \omega_{23} \Delta t_{23})} \right]\;. \nn
\eea
This expression contains quantum mechanical off-shell and memory 
effects. However, it does not yet contain corrections from 
interactions with SM gauge bosons, which introduce thermal damping 
and the associated widths
for the Higgs and lepton fields. As explained in a detailed discussion
in \cite{Anisimov:2010dk}, this fact is responsible for the apparent difference
between quantum Boltzmann solutions and the present solution of Kadanoff-Baym 
equations. Moreover, for $T \gtrsim M$ these damping widths are of order 
$\gamma_{l,\phi}\sim g_\mathrm{SM}^2 T \gg \lambda^2 T \sim \Gamma$, i.e.~larger than the heavy
neutrino rate, and hence can modify the result 
significantly \cite{Anisimov:2010aq}. Any quantitative theory of quantum leptogenesis 
therefore requires a systematic inclusion of these effects.

\section{Gauge corrections to leptogenesis}
\label{sec:gauge}

The inclusion of gauge corrections to thermal leptogenesis is complicated
by the fact that it happens before the electroweak transition, i.e.\ gauge bosons
are massless and the non-abelian weak dynamics are QCD-like \cite{Bergerhoff:1994sj,Philipsen:1996af,Philipsen:1997rq}. This in particular implies
the dynamical generation of mass scales $\sim T, g T, g^2T$, of which the ultra-soft
$\sim g^2T$ is entirely non-perturbative. For processes with 
gauge particles on the soft scale $\sim gT$,
hard thermal loops (HTL) contribute the same power in the coupling constant to the self-energy to any loop order, and hence have to be resummed when counting 
orders in the gauge couplings \cite{Bellac:2011kqa,Kapusta:2006pm}.
Here we consider fermion processes, 
whose external momenta in a plasma are typically hard, $k\sim T$.
Nevertheless, for hard momenta near the light cone, $k^2\sim g_\mathrm{SM}^2T^2$, 
loop momenta collinear with the external ones require a similar resummation of collinear
thermal loops (CTL) \cite{Besak:2010fb}. For power counting purposes, all SM couplings are considered
as parametrically the same.

\subsection{CTL resummation of the Majorana neutrino self-energy}
\label{sec:CTLResummation}

\begin{figure}[t]
	\begin{center}
		\includegraphics[width=0.9\textwidth]{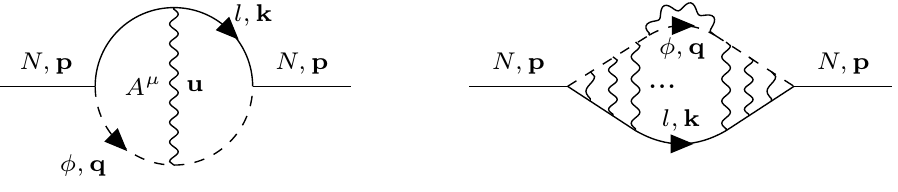}
	\end{center}
	\caption[]{Majorana neutrino self-energy contributions with one and several 
	additional gauge bosons with soft three-momentum $\mathbf{u}\sim g_\mathrm{SM}T$.}
	\label{fig:Correction}
\end{figure}

As a first step, we need the gauge corrections to the Majorana neutrino self-energy and its proper resummation shown in figure~\ref{fig:Correction}
to determine the Majorana neutrino production rate \cite{Anisimov:2010gy}.
One effect of interactions with the plasma is to modify the dispersion relations of
the thermalised leptons and Higgs particles by the asymptotic thermal masses\footnote{Since other
Yukawa couplings are very small, we include only the coupling of the top quark.} 

\bea
m_{\phi, \infty} &=& \frac{1}{16} (3g^2 + g'^2 + 4h_t^2 + 8 \lambda_{\phi})T^2\;, \nn\\ 
m_{l, \infty} &=& \frac{1}{16} (3 g^2 + g'^2) T^2\;.
\label{eq:m_as}
\eea
Next, we consider nearly collinear momenta $k$ close to the light cone, pointing
approximately in the three-direction of 
the light-like four-vector $v = (1, \mathbf{v}), \mathbf{v}^2 = 1$. 
Three-momenta are then specified by 
components parallel and perpendicular to $\mathbf{v}$,
\begin{align}
	\mathbf{k}=k_\parallel\,\mathbf{\hat{v}}+\mathbf{k}_\perp\;,\quad \mbox{with}\quad
	k_{\parallel} := \mathbf{k} \cdot \mathbf{v}\;,\quad
	\mathbf{k}_{\perp}\cdot\mathbf{v}=0\;.
\end{align}
The leptonic and Higgs loop momenta thus satisfy
\begin{align}
k_\parallel \sim T\;,\quad |\mathbf{k}_{\perp}| \sim g_\mathrm{SM} T\;,\quad v\cdot k\sim g_\mathrm{SM}^2T\;,\quad
	k^2\sim g_\mathrm{SM}^2T^2\;,
\end{align}
and thermal masses are important. Further, the gauge loop momenta are considered soft, 
$u_\mu\sim g_\mathrm{SM}T$.

In this kinematical region the diagrams without and with gauge corrections,
figures~\ref{fig:MajoranaSelf} and \ref{fig:Correction}, respectively,
are parametrically the same,
\bea
\Sigma_{\text{1-loop}}& \sim &\underbrace{g_\mathrm{SM}^2}_{\text{3-vertex factor}} \;
\underbrace{\left( \frac{1}{g_\mathrm{SM}^2}\right)^2}_{\text{propagators}} 
\underbrace{g_\mathrm{SM}^4}_{\text{phase space integration}} \sim g_\mathrm{SM}^2\;,\nn\\
\Sigma_{\text{2-loop}}&\sim & \underbrace{g_\mathrm{SM}^2}_{\text{3-vertex factor}} \;
\underbrace{g_\mathrm{SM}^2}_{\text{gauge boson vertices}} 
\underbrace{\left( \frac{1}{g_\mathrm{SM}^2} \right)^5}_{\text{propagators}} 
\underbrace{(g_\mathrm{SM}^4)^2}_{\text{phase space integration}} \sim g_\mathrm{SM}^2\;.
\eea
The 3-vertex factor in each expression arises kinematically, where 
we followed the power counting rules from~\cite{Besak:2010fb,Anisimov:2010gy}. 
Thus, adding soft internal gauge boson lines does not increase the
order in the gauge coupling and necessitates a resummation of the entire ladder to infinite loop order,
figure~\ref{fig:Correction},  while contributions from crossed ladder rungs and vertex corrections to loop particles
are further suppressed \cite{Besak:2010fb,Anisimov:2010gy}. 

The problem corresponds to the Landau-Pomeranchuk-Migdal effect and has also been solved for gluon radiation in a QCD plasma \cite{Arnold:2002ja}. 
Its application to the heavy Majorana neutrino
was treated in \cite{Anisimov:2010gy}, where the ladder-resummed self-energy, including thermals widths of the Higgs and leptons, is written as
\begin{align}
\label{eq:corrSelf}
\Sigma_{R, \bp}^\mathrm{ret}(\omega) = -\lambda^2 \frac{\text{i}}{2} \int_{\mathbf{k}} \frac{d(r) \mathcal{F}(p_{\parallel}, k_{\parallel})}{k_{\parallel}-p_{\parallel}} \begin{pmatrix}
-\frac{k_1 - \text{i} k_2}{2 k_{\parallel}} \\
1
\end{pmatrix} \cdot \begin{pmatrix}
-\frac{f^1 + \text{i}f^2}{4 k_{\parallel}}\;, && \psi
\end{pmatrix}
\end{align}   
with $d(r) = 2$ the dimension of the $SU(2)$ fundamental representation and the combination of Fermi-Dirac and Bose-Einstein distributions
\begin{align}
\mathcal{F}(p_{\parallel}, k_{\parallel}) = f_F(k_{\parallel}) + f_B(k_{\parallel}-p_{\parallel})\;.
\end{align} 
The functions $\psi$ and $\mathbf{f} = (f^1, f^2)$ are defined as solutions of the integral equations
\begin{align}
\label{eq:intEq1}
\text{i}\epsilon(p, \mathbf{k}) \mathbf{f}(\mathbf{k}_{\perp}) - \int \frac{\diff^2 u_{\perp}}{(2 \pi)^2} \mathcal{C}(|\mathbf{u}_{\perp}|) \left[ \mathbf{f}(\mathbf{k}_{\perp}) - \mathbf{f}(\mathbf{k}_{\perp}-\mathbf{u}_{\perp}) \right] &= 2 \mathbf{k}_{\perp}\;, \\
\label{eq:intEq2}
\text{i} \epsilon(p, \mathbf{k}) \psi(\mathbf{k}_{\perp}) - \int \frac{\diff^2 u_{\perp}}{(2 \pi)^2} \mathcal{C}(|\mathbf{u}_{\perp}|) \left[ \psi(\mathbf{k}_{\perp}) - \psi(\mathbf{k}_{\perp}- \mathbf{u}_{\perp}) \right] &= 1\;,
\end{align}
with the kernel
\begin{align}
\mathcal{C}(|\mathbf{u}_{\perp}|) = T \left[ C_2(r)g^2 \left( \frac{1}{|\mathbf{u}_{\perp}|^{2}}- \frac{1}{|\mathbf{u}_{\perp}|^{2}+m_D^2}\right) + y_l^2 g'^2 \left( \frac{1}{|\mathbf{u}_{\perp}|^{2}}- \frac{1}{|\mathbf{u}_{\perp}|^{2}+m_D'^2} \right) \right]\;.
\end{align}
The kernel is obtained from the gauge field propagator and contains the ladder resummation, 
with the $SU(2)$ Casimir operator $C_2(r) = 3/4$, the hypercharge $y_l = -1/2$, and the HTL-resummed Debye 
masses \cite{Carrington:1991hz}
\begin{align}
m_D^2 = \frac{11}{6} g^2 T^2\;,~~~~~~m_D^{\prime 2} = \frac{11}{6} g^{\prime 2} T^2\;.
\end{align}
The quantity $\epsilon(p,\mathbf{k})$ is given in a frame, where $\mathbf{v} \parallel \bp$ and $\bp_{\perp} = 0$, as
\begin{align}
\epsilon(\omega, p_{\parallel}, \mathbf{k}) = \alpha( \omega, p_{\parallel}, \mathbf{k}) + \beta(p_{\parallel}, k_{\parallel}) \mathbf{k}_{\perp}^2 = \beta(p_{\parallel}, k_{\parallel}) (M_{\text{eff}}^2 + \mathbf{k}_{\perp}^2)\;,
\end{align}
with
\begin{align}
\alpha( \omega, p_{\parallel}, k_{\parallel}) &= \omega - p_\parallel + \frac{m_{\phi, \infty}^2}{2 (k_\parallel - p_\parallel)} - \frac{m_{l, \infty}^2}{2 k_\parallel}\;, \\
\beta(p_{\parallel}, k_{\parallel}) &= \frac{p_{\parallel}}{2 k_{\parallel}(k_{\parallel} - p_{\parallel})}\;, \\
M_{\text{eff}}^2 &= \frac{m_{l, \infty}^2 (p_{\parallel} - k_{\parallel}) + m_{\phi, \infty}^2 k_{\parallel} + 2 k_{\parallel}(\omega - p_{\parallel})(k_{\parallel} - p_{\parallel})}{p_{\parallel}}\;.
\end{align}
The integral eqs.~\eqref{eq:intEq1}, \eqref{eq:intEq2} are best solved in  
Fourier space, 
\begin{align}
\psi(\mathbf{b}) = \int \frac{\diff^2 k_{\perp}}{(2 \pi)^2}  \psi(\mathbf{k}_{\perp}) \text{e}^{\text{i} \mathbf{k}_{\perp} \mathbf{b}},~~~~~\mathbf{f}(\mathbf{b}) = \int \frac{\diff^2 k_{\perp}}{(2 \pi)^2}  \mathbf{f}(\mathbf{k}_{\perp}) \text{e}^{\text{i} \mathbf{k}_{\perp} \mathbf{b}}\;. 
\end{align}
Rotational invariance implies $\psi(\mathbf{b}) = \psi(b), \mathbf{f}(\mathbf{b}) = h(b) \mathbf{b}$, and the equations are converted into ordinary
differential equations,
\begin{align}
\label{eq:ODE1}
-\text{i} \beta \left( \partial_b^2 + \frac{1}{b} \partial_b - M_{\text{eff}}^2\right) \psi(b) - \mathcal{K}(b) \psi(b) &= 0\;, \\ 
\label{eq:ODE2}
-\text{i} \beta \left( \partial_b^2 + \frac{3}{b} \partial_b - M_{\text{eff}}^2\right) h(b) - \mathcal{K}(b) h(b) &= 0\;,
\end{align}
with $\mathcal{K}(b)$ the Fourier transform of $\mathcal{C}(\mathbf{u}_{\perp})$\footnote{The calculation of $\mathcal{K}(b)$ requires a renormalization procedure to perform the Fourier transformation as well as the $\mathbf{u}_{\perp}$ integral. A detailed calculation can be found in \cite{Anisimov:2010gy}.} 
\begin{align}
\mathcal{K}(b) = \frac{1}{(2 \pi)^2} \int_0^{2 \pi} \diff \phi \int_0^\infty \diff |\mathbf{u}_{\perp}| \left( 1 - \text{e}^{\text{i} |\mathbf{u}_{\perp}| b \cos (\phi)} \right) |\mathbf{u}_{\perp}| \, \mathcal{C}(|\mathbf{u}_{\perp}|) \;.
\end{align}
The solution of eqs.~\eqref{eq:ODE1}, \eqref{eq:ODE2} determines the Majorana neutrino self-energy in the form
\begin{align}
\label{eq:corrSelfResult}
\text{Im} \left( \Sigma_{R, \bp}^\mathrm{ret} (\omega) \right) &= -\frac{\lambda^2 d(r)}{2} \int \frac{\diff k_{\parallel}}{2 \pi} \frac{\mathcal{F}(p_{\parallel}, k_{\parallel})}{k_{\parallel}-p_{\parallel}} \begin{pmatrix}
\frac{\text{Im}(c_{2,h}(\omega, p_{\parallel}, k_{\parallel}))}{4 k_{\parallel}^2} && 0 \notag \\
0 && \text{Re}(c_{2,\psi}(\omega, p_{\parallel}, k_{\parallel}))
\end{pmatrix} \\ &\equiv -\frac{\lambda^2 d(r)}{2} \begin{pmatrix}
\sigma_{h}(\omega, p_{\parallel}) && 0 \\
0 && \sigma_{\psi}(\omega, p_{\parallel})
\end{pmatrix}\;,
\end{align}
where the  $c_2$ are asymptotic solutions
\begin{align}
\int \frac{\diff^2 k_{\perp}}{(2 \pi)^2} \text{Re}(\mathbf{k}_{\perp} \cdot \mathbf{f}_{\perp}(\mathbf{k}_{\perp})) &= \lim_{b \to 0} \text{Im}(2 h(b)) \equiv 2 \text{Im}(c_{2,h}(\omega, p_{\parallel}, k_{\parallel}))\;, \\
\int \frac{\diff^2k_{\perp}}{(2 \pi)^2} \text{Re}(\psi(\mathbf{k}_{\perp})) 
	&= \lim_{b \to 0} \text{Re}(\psi(b)) \equiv 
	\text{Re}(c_{2,\psi}(\omega, p_{\parallel}, k_{\parallel}))\;.
\end{align}
The off-diagonal elements of eq.~\eqref{eq:corrSelfResult} receive contributions vanishing in the 
limit $b \to 0$, and $\mathcal{O}(1/b)$ divergent contributions. 
These are removed by renormalization, since they appear in the temperature 
independent part of the self-energy, while the temperature dependent part is given by the function $\mathcal{F}(p_{\parallel}, k_{\parallel})$. 
For later use in the lepton number density we define the functions
\begin{align}
\label{eq:sigmah}
\sigma_{h}(\omega, p_{\parallel}) &\equiv \int \frac{\diff k_{\parallel}}{2 \pi} \frac{\mathcal{F}(p_{\parallel}, k_{\parallel})}{4 k_{\parallel}^2(k_{\parallel}-p_{\parallel})} \text{Im}(c_{2,h}(\omega, p_{\parallel}, k_{\parallel}))\;, \\ 
\label{eq:sigmapsi}
\sigma_{\psi}(\omega, p_{\parallel}) &\equiv \int \frac{\diff k_{\parallel}}{2 \pi} \frac{\mathcal{F}(p_{\parallel}, k_{\parallel})}{k_{\parallel}-p_{\parallel}} \text{Re}(c_{2,\psi}(\omega, p_{\parallel}, k_{\parallel}))\;. 
\end{align}
Because of the symmetry between right- and left-handed Dirac components, we have 
\bea
\label{eq:selfSymm}
\Sigma_{R, \bp}^\mathrm{ret} (\omega)& = &\Sigma_{L, -\bp}^\mathrm{ret}(\omega)\;, \\
\label{eq:sigmaSymm}
\sigma_h(-\omega, p_{\parallel}) = \sigma_{\psi}(\omega,p_{\parallel})\;&, &\quad \sigma_h(\omega, - p_\parallel) = \sigma_{\psi}(\omega,p_{\parallel})\;.
\eea

\begin{figure}[t]
\centerline{
\includegraphics[width=0.5\textwidth]{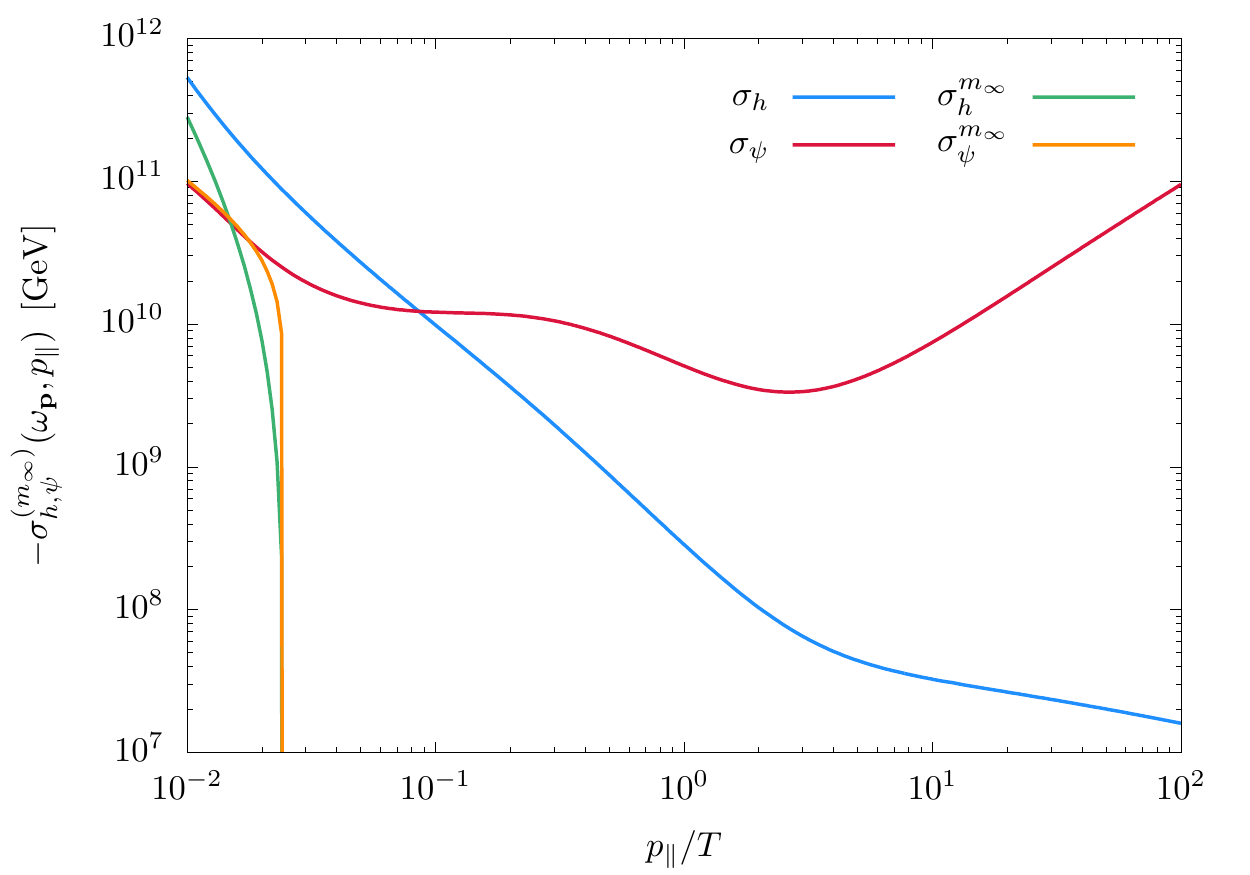}		
}	
\caption[]{The fully gauge corrected $\sigma_{h,\psi}(\omega_{\mathbf{p}}, p_{\parallel})$ and the $\sigma^{m_{\infty}}_{h,\psi}(\omega_{\mathbf{p}}, p_{\parallel})$, including asymptotic masses only, evaluated on-shell, $\omega=\omega_{\mathbf{p}}$, with $M=10^{10} \, \mathrm{GeV}$ and $T=10^{11} \, \mathrm{GeV}$.}
\label{fig:sigmaPlot}
\end{figure}
In order to quantify the effect of the ladder resummation, it is useful to also analyse eqs.~\eqref{eq:intEq1}, \eqref{eq:intEq2} 
with $\mathcal{C}(|\mathbf{u}_\perp|) = 0$, giving a Majorana neutrino self-energy which is only partly corrected by asymptotic 
masses. 
The equations in this limit can be solved trivially and one obtains after $\bf k_{\perp}$-integration
\begin{align}
\label{eq:sigmahtree}
\sigma_h^{m_{\infty}}(\omega, p_{\parallel}) =& \int_{k_{\text{low}}}^{k_{\text{high}}} \frac{dk_{\parallel}}{2 \pi} \frac{\mathcal{F}(p_{\parallel}, k_{\parallel})}{k_{\parallel}-p_{\parallel}} \frac{\alpha(\omega, p_{\parallel}, k_{\parallel})}{16 k_{\parallel}^2 \beta(p_{\parallel}, k_{\parallel})|\beta(p_{\parallel}, k_{\parallel})|}\;, \\
\label{eq:sigmapsitree}
\sigma_{\psi}^{m_{\infty}}(\omega, p_{\parallel}) =& -\int_{k_{\text{low}}}^{k_{\text{high}}} \frac{dk_{\parallel}}{2 \pi} \frac{\mathcal{F}(p_{\parallel}, k_{\parallel})}{k_{\parallel}-p_{\parallel}} \frac{1}{4|\beta(p_{\parallel}, k_{\parallel})|}\;.
\end{align}
The integration boundaries in this case are given by
\begin{align}
k_{\text{low}} = \frac{X - \sqrt{Y}}{4(\omega - p_{\parallel})}\;,~~~~~k_{\text{high}} = \frac{X + \sqrt{Y}}{4(\omega - p_{\parallel})}\;,
\end{align}
with
\beq
X \equiv m_{l,\infty}^2 - m_{\phi,\infty}^2 + 2(\omega-p_{\parallel})p_{\parallel}\;,\quad
Y \equiv X^2 - 8 m_{l,\infty}^2(\omega - p_{\parallel})p_{\parallel}\;,\quad
Y \geq 0\;.
\eeq 

In figure \ref{fig:sigmaPlot} the fully corrected $\sigma_{h,\psi}(\omega_{\mathbf{p}},p_{\parallel})$ are plotted 
for $M=10^{10} \, \mathrm{GeV}$, $T=10^{11} \, \mathrm{GeV}$, and compared to the $\sigma^{m_{\infty}}_{h,\psi}(\omega_{\mathbf{p}}, p_{\parallel})$.
The latter are significantly smaller than their fully resummed counterparts in most of the parameter space considered. Note that above a 
certain $p_{\parallel}$, the $\sigma^{m_{\infty}}_{h,\psi}(\omega_{\mathbf{p}},p_{\parallel})$ vanish kinematically. 

For the SM corrected Majorana decay width we start with eq.~\eqref{eq:ConnDecaySelf}. In Weyl basis it is possible to simplify further, 
\begin{align}
\Gamma_{\bp}(\omega) &= \frac{1}{2 \omega} \text{tr}\left[ \slashed p \text{Im} \left( \Sigma^\mathrm{ret}_{\bp}(\omega) \right) \right] \\
&= \frac{1}{2 \omega} \text{tr} \left[ \begin{pmatrix}
0 && \sigma \cdot p \\
\bar{\sigma} \cdot p && 0 \\
\end{pmatrix} \cdot \begin{pmatrix}
0 && \text{Im}\left( \Sigma^\mathrm{ret}_{R, \bp}(\omega) \right) \\
\text{Im}\left( \Sigma^\mathrm{ret}_{L, \bp}(\omega) \right) && 0
\end{pmatrix} \right]\;.
\end{align} 
Using $\bp_{\perp} = 0$ we find
\begin{align}
\Gamma_{\bp}(\omega) = - \frac{\lambda^2 d(r)}{2 \omega} \left( (\omega + p_{\parallel}) \sigma_h(\omega, p_{\parallel}) 
+ (\omega - p_{\parallel}) \sigma_{\psi}(\omega, p_{\parallel}) \right)\;.
\end{align}
For all further calculations we only need the on-shell expression $\Gamma_{\bp} = \Gamma_{\bp} (\omega_{\bp})$ 
in the Majorana neutrino propagator, see eq.~\eqref{eq:MajProp}. Note that the symmetry property for the $\sigma_{h,\psi}$-functions
 \eqref{eq:sigmaSymm} reflects the symmetry of $\Gamma_\bp$, eq.~\eqref{eq:gammaSym}. 

\subsection{Gauge corrections to lepton self-energies}
\label{sec:cylinder}

We now have everything at hand to also   
include gauge corrections to the $CP$-violating diagrams in figure~\ref{fig:CpViolation}, which enter
the lepton number matrix. Adding one internal gauge boson line modifies those diagrams by either
self-energy or vertex corrections of the internal lines, the resulting three-loop diagrams  are collected
in appendix \ref{sec:app_a}. From the last section it is clear that in the collinear kinematical region
this will not suffice for a result to leading order in the gauge coupling. Instead, infinitely many
insertions have to be resummed.

\begin{figure}[t]
\centerline{
\includegraphics[width=0.8\textwidth]{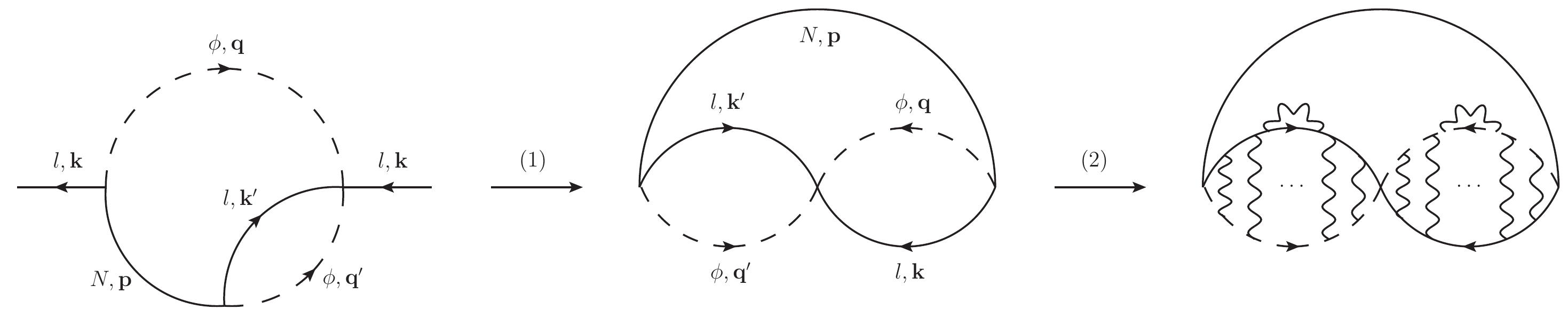}
}
\caption[]{Transformation of the lepton self-energy diagram to a form amenable to resummation.}
\label{fig:convert_diag}
\end{figure}
In order to achieve this, we follow the procedure depicted in figure~\ref{fig:convert_diag} and first close the external fermion lines (1), 
which implies an integration over lepton 
momenta and transforms the lepton number matrix into the lepton number density,
\beq
\int \frac{\diff^3k}{(2\pi)^3} \,L_{\mathbf{k},ii} =n_{L,ii}\;.
\eeq
In this form of the diagram, 
we can focus on the neutrino self-energy and apply the resummation from section \ref{sec:CTLResummation}.
The only additional feature is the effective four-point vertex appearing in this self-energy, so that two of the resummed
blobs from   \ref{sec:CTLResummation} are needed in step (2).
It is instructive to 
check by direct comparison with appendix \ref{sec:app_a} that indeed all parametrically leading three-loop diagrams shown there 
are included and resummed. In appendix \ref{sec:app_resum} we give a more detailed derivation of the double-blob diagram, which
is general and also applies away from the hierarchical mass scenario.

\subsection{Lepton number matrix with complete SM corrections}

Collecting all results from the previous sections it is possible to give an expression for the SM corrected lepton number matrix. Starting with eq.~\eqref{eq:TreeLeptonNumberMatrix} we first carry out the $\mathbf{q}$ and $\mathbf{q}'$ momentum integrals by using the momentum conserving $\delta$-functions, so that the trace in the integrand takes the form
\bea
\hspace{-7mm} \int_{\mathbf{k}, \mathbf{k}'} \text{tr} \left( \text{Im}\Sigma_{R, \mathbf{k}, -\bp-\mathbf{k}}^\mathrm{ret}(\omega_{21}) \text{Im}\Sigma_{L, \mathbf{k}', \bp - \mathbf{k}'}^\mathrm{ret}(\omega_{23}) \right) &=& \text{tr} \left( \text{Im}\Sigma_{R, -\bp}^\mathrm{ret}(\omega_{21}) \text{Im}\Sigma_{L, \bp}^\mathrm{ret}(\omega_{23}) \right)\nn\\
&=&\text{tr} \left( \text{Im}\Sigma_{R, \bp}^\mathrm{ret}(\omega_{21}) \text{Im}\Sigma_{R, \bp}^\mathrm{ret}(\omega_{23}) \right)\;,
\eea  
where we have used the relation between the left- and right-handed self-energy \eqref{eq:selfSymm}.

Finally we use the resummed diagram figure \ref{fig:convert_diag} to insert the resummed Majorana neutrino self-energy in the equation above.
Then, the trace can be carried out using eq.~\eqref{eq:corrSelfResult},
\begin{align}
\text{tr} \left( \text{Im}\Sigma_{R, \bp}^\mathrm{ret}(\omega_{21}) \text{Im}\Sigma_{R, \bp}^\mathrm{ret}(\omega_{23}) \right) = \sigma_{\psi}(\omega_{21}, p_{\parallel})\sigma_{\psi}(\omega_{23}, p_{\parallel}) + \sigma_h(\omega_{21}, p_{\parallel})\sigma_h(\omega_{23}, p_{\parallel})\;. 
\end{align}
Altogether, we arrive at the following result of the corrected lepton number density:
\begin{align}
\label{eq:final}
n&_{L,ii}(t) = -\frac{128}{\pi} \epsilon_{ii} \int\limits_0^t \diff t_1 \int\limits_0^t \diff t_2 \int\limits_0^{t_2} \diff t_3 \int\limits_{0}^{\infty} \diff p_\parallel \int\limits_{-\infty}^{\infty} \frac{\diff \omega_{21}}{2 \pi} \int\limits_{-\infty}^{\infty} \frac{\diff \omega_{23}}{2 \pi} \frac{p_{\parallel}^2}{\omega_{\bp}} f_F(\omega_{\bp})f_F(\omega_{21}) \cos(\omega_{\bp} \Delta t_{13}) \notag \\ 
	&\times \text{e}^{-\Gamma_{\bp} \frac{t_{13}}{2}} \text{Re} \left( \text{e}^{-\text{i}(\omega_{21}\Delta t_{21} + \omega_{23}\Delta t_{23})} \right) [\sigma_{\psi}(\omega_{21}, p_\parallel)\sigma_{\psi}(\omega_{23}, p_\parallel) + \sigma_h(\omega_{21}, p_\parallel)\sigma_h(\omega_{23}, p_\parallel)]\;.
\end{align} 
This expression contains all SM corrections to leading order in the respective couplings and for the first time gives a complete
description of quantum leptogenesis.

\section{Evaluation of the lepton number density}
\label{sec:num}

It now remains to numerically evaluate our final result
eq.~\eqref{eq:final}. As we shall see shortly, the time integrations can be performed
analytically. However, the remaining three-dimensional 
integral has an integrand with rapidly oscillating parts, 
rendering an accurate numerical evaluation difficult.
Moreover, $\sigma_h$ and 
$\sigma_{\psi}$ are specified by the asymptotic solutions of the differential
eqs.~\eqref{eq:ODE1} and \eqref{eq:ODE2}, which have to be solved numerically for each call of those functions. 
A brute force numerical result would thus 
require a significant amount of computing power.

Here we adopt a different approach. Since the final result is 
accurate to leading order in the couplings only, we limit the integration range
to those regions, which give the parametrically leading contribution. This permits
us to do two more integrations analytically, leaving only the $p_\parallel$-integral
as a final numerical task. Since we are left with only one momentum variable, we simplify the notation 
$p_\parallel\to p$ in this section.

\subsection{The time integrals}
\label{sec:t_int}

We begin by factoring out the time integrals,
\bea
\label{eq:t-int}
n_{L,ii}(t)=-\frac{128}{\pi}\epsilon_{ii}
 \int\limits_{0}^{\infty} \diff p \int\limits_{-\infty}^{\infty} 
 \frac{\diff \omega_{21}}{2 \pi} \int\limits_{-\infty}^{\infty} 
 \frac{\diff \omega_{23}}{2 \pi}\;&&
\mathcal{T}(t;\omega_{21},\omega_{23},p)
\frac{p^2}{\omega_\bp}f_F(\omega_\bp)f_F(\omega_{21})\\
&&\times [\sigma_{\psi}(\omega_{21}, p)\sigma_{\psi}(\omega_{23}, p) 
+ \sigma_h(\omega_{21}, p)\sigma_h(\omega_{23}, p)]\;. \nn
\eea
The resulting
function, 
\beq
\label{eq:timeDepPart}
\mathcal{T}(t;\omega_{21},\omega_{23},p)\equiv 
\int\limits_0^t \diff t_1 \int\limits_0^t \diff t_2 \int\limits_0^{t_2} 
\diff t_3 \cos(\omega_{\bp} \Delta t_{13}) \text{e}^{-\Gamma_{\bp}\frac{t_{13}}{2}} 
\text{Re} \left( \text{e}^{-\text{i}(\omega_{21} \Delta t_{21} 
+ \omega_{23} \Delta t_{23})} \right)\;, 
\eeq
can be integrated explicitly with the help of \texttt{Mathematica},
and the result is spelled out in appendix \ref{sec:app_b}.
We now use this expression 
to estimate in which domain of frequencies it gives
parametrically dominant contributions to the integral.

The denominator of \eqref{eq:Tfactor} reads
\begin{align}
\label{eq:denominator}
(\omega_{21} + \omega_{23})&(\Gamma_{\bp}^2 + 4(\omega_{21} - \omega_{\bp})^2)(\Gamma_{\bp}^2+4(\omega_{21} + \omega_{\bp})^2) \nonumber \\
\times &(\Gamma_{\bp}^2+4(\omega_{23} - \omega_{\bp})^2)(\Gamma_{\bp}^2+4(\omega_{23} 
	+ \omega_{\bp})^2)\;.
\end{align}
At typical particle momentum in our system $p \sim \mathcal{O}(T)$, the Majorana neutrino decay width is of order $\Gamma_{\bp} \lesssim \mathcal{O}(\lambda^2 T)$, and since we are working at temperatures $T \gtrsim M$, this results in $\omega_{\bp} \sim \mathcal{O}(T)$ as well.
Hence in the four factors including $\Gamma_{\bp}$, the latter is subdominant compared
to the frequency terms. These factors are thus smallest when the two frequencies 
approach the mass shell. Alternatively, there is a pole due to the first factor. 
The function $\mathcal{T}$ thus has peaks in the following two limits, 
for which we estimate their parametric contribution:

\begin{enumerate}
	\item $\omega_{21},\omega_{23}\to \pm \omega_{\bp}$: symmetric under $\omega_{\bp} \to -\omega_{\bp}$, 
	\begin{align}
	\lim_{\omega_{23},\omega_{21} \to \pm\omega_{\bp}} \mathcal{T}(t; \omega_{21}, \omega_{23}, p) 
		=& \frac{2 \text{e}^{-\Gamma_{\bp}t}(\text{e}^{\Gamma_{\bp}t/2}-1) }{\Gamma_{\bp}\omega_{\bp}(\Gamma_{\bp}^2 + 16 \omega_{\bp}^2)} \nn \\ \nn &\times
		\left[ 4 \omega_{\bp} + \text{e}^{\Gamma_{\bp} t/2}(-4\omega_{\bp}\cos(\omega_{\bp} t) + \Gamma_{\bp}\sin(2 \omega_{\bp}t)) \right]\nn\\
		 \sim & \mathcal{O}(\lambda^{-2} T^{-3})\;.
	\end{align}
	\item  $\omega_{21},-\omega_{23}\to \pm \omega_{\bp}$: symmetric under $\omega_{\bp} \to -\omega_{\bp}$,
	\bea
	\label{eq:TDomPeak}
	\lim_{-\omega_{23},\omega_{21} \to \pm\omega_{\bp}} \mathcal{T}(t; \omega_{21}, \omega_{23}, p) 
		&=& \frac{\text{e}^{-\Gamma_{\bp}t}}{\Gamma_{\bp}^3 (\Gamma_{\bp}^2 
		+ 16 \omega_{\bp}^2)^2} \bigg[ -8(\Gamma_{\bp}^4 
		+ 16 \Gamma_{\bp}^2 \omega_{\bp}^2 +128 \omega_{\bp}^4) \notag \\ \notag 
		&&+ 4 \text{e}^{\Gamma_{\bp}t}\Big(\Gamma_{\bp}^4(\Gamma_{\bp} t -2) +8 (3 \Gamma_{\bp} t - 4) \Gamma_{\bp}^2\omega_{\bp}^2+128(\Gamma_{\bp} t -2) \omega_{\bp}^4\Big) \\
		&&-2 \text{e}^{\Gamma_{\bp}t/2}\Big((\Gamma_{\bp} t -4) 
		(\Gamma_{\bp}^2 + 16 \omega_{\bp}^2)^2 \nn\\ 
		&&+ \Gamma_{\bp}^3(\Gamma_{\bp}(\Gamma_{\bp}t -4) 
		+16 \omega_{\bp}^2t) \cos(2 \omega_{\bp} t) \Big) \bigg]\nn\\
		&	\sim& \mathcal{O}(\lambda^{-6} T^{-3})\;.
	\eea
\end{enumerate}
\begin{figure}[t]
\vspace*{-2cm}
	\centerline{
	\includegraphics[width=0.8\textwidth]{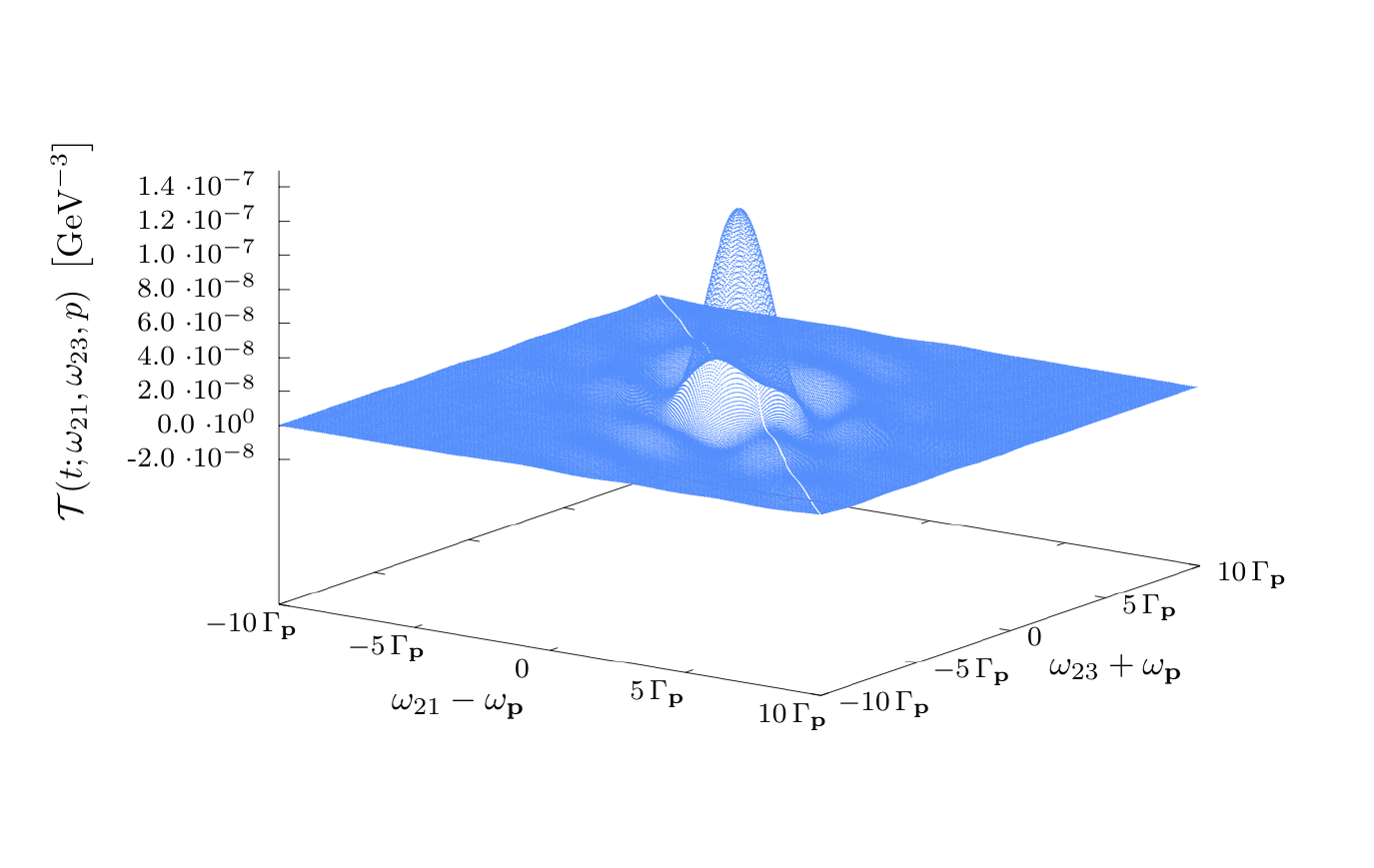}
	}
	\vspace*{-1.0cm}
\caption{$\mathcal{T}(t; \omega_{21}, \omega_{23}, p)$ around $\omega_{21} = -\omega_{23} = \omega_{\bp}$ for fixed $t = 10^{-2}/\mathrm{GeV}$ and $\omega_{\bp} =1.42 \cdot 10^{10} \, \mathrm{GeV}$ implying $\Gamma_\bp = 2.01 \cdot 10^2 \, \mathrm{GeV}$ at $T = 10^{11} \, \mathrm{GeV}$, $M = 10^{10} \, \mathrm{GeV}$, and $\lambda^2 = 10^{-8}$.} 
\label{fig:data1}
\end{figure}
Sufficiently far away from the peaks we can simplify $\mathcal{T}$ by
considering $\Gamma_\bp \to 0$, while keeping $\Gamma_{\bp}t =$ const., 
which is possible because of the smallness of $\Gamma_{\bp}$ \cite{Anisimov:2010dk},
\begin{align}
\label{eq:ToffRegion}
\lim_{\Gamma_{\bp} \to 0} \mathcal{T}(t; \omega_{21}, \omega_{23}, p) = &\frac{\text{e}^{-\Gamma_\bp t /2}}{(\omega_{21} + \omega_{23}) (\omega_{\bp}^2 - \omega_{23}^2) (\omega_{\bp}^2 - \omega_{21}^2)}  \nonumber \\
&\times \left[ (\omega_{21} - \omega_{23}) \omega_{\bp} \sin (t \omega_{\bp})  \right.(\cos (t \omega_{21}) - \cos (t \omega_{23})) \nn \\ 
&+ (\omega_{21} \omega_{23} - \omega_{\bp}^2) (\cos (t \omega_{\bp}) (\sin (t \omega_{21}) + \sin (t \omega_{23}))  \nonumber \\
& - \left. \text{e}^{\Gamma_\bp t /2} \sin (t (\omega_{21} + \omega_{23}))) \right]\;.
\end{align}
Since this is symmetric under the exchange of $\omega_{21}$ and $\omega_{23}$, 
we now consider $|\omega_{21}| > |\omega_{23}|$ without  
loss of generality. 
An upper bound is obtained by approximating 
$(\omega_{21} + \omega_{23}) \approx \omega_{21}$ in the denominator. 
We then have three cases of interest:
\begin{enumerate}
\item $\omega_{\bp} \ll |\omega_{21}|, |\omega_{23}|$: this is far away from the peaks and we have $(\omega_{\bp}^2 - \omega_{21}^2) \approx -\omega_{21}^2$, 
	$(\omega_{\bp}^2 - \omega_{23}^2) \approx -\omega_{23}^2$, 
and $(\omega_{21} \omega_{23} - \omega_{\bp}^2) \approx \omega_{21}\omega_{23}$. 
\beq
 \mathcal{T}(t; \omega_{21}, \omega_{23}, p) \sim \mathcal{O}(\omega_{21}^{-2} \omega_{23}^{-1})\;.
\eeq
\item $\omega_{21} \gg \omega_{\bp}$ but $\omega_{23} \ll \omega_{\bp}$: in this case 
$(\omega_{\bp}^2 - \omega_{21}^2) \approx -\omega_{21}^2$, 
$(\omega_{\bp}^2 - \omega_{23}^2) \approx \omega_{\bp}^2$. 
\beq
 \mathcal{T}(t; \omega_{21}, \omega_{23}, p) \sim \mathcal{O}(\omega_{21}^{-2} T^{-1})\;.
\eeq
\item $\omega_{\bp} \gg |\omega_{21}|, |\omega_{23}|$: here we approximate $(\omega_{\bp}^2 - \omega_{21}^2) \approx \omega_{\bp}^2$, $(\omega_{\bp}^2 - \omega_{23}^2) \approx \omega_{\bp}^2$, and $(\omega_{21} \omega_{23} - \omega_{\bp}^2) \approx -\omega_{\bp}^2$. 
\beq
 \mathcal{T}(t; \omega_{21}, \omega_{23}, p) \sim \mathcal{O}(\omega_{21}^{-1} T^{-2})\;.
\eeq
For very small values of $|\omega_{21}|$ we may expand the trigonometric functions
in eq.~\eqref{eq:ToffRegion}, leading to
\begin{align}
 \mathcal{T}(t; \omega_{21}, \omega_{23}, p) \sim \frac{\Gamma_\bp t \left(1- \text{e}^{-\Gamma_\bp t/2} \cos(\omega_{\bp}t) \right)}{\Gamma_\bp \omega_{\bp}^2}\;,
\end{align}
where we have kept $\Gamma_\bp$ to compare with the peak region.
For $\Gamma_\bp t \sim \mathcal{O} (1)$ one finds 
\beq
 \mathcal{T}(t; \omega_{21}, \omega_{23}, p) \sim \mathcal{O} (\lambda^{-2} T^{-3})\;.
\eeq
\end{enumerate}
Comparing all estimates shows that the region around the peak at 
$\omega_{21} = -\omega_{23} = \pm \omega_{\bp}$ gives the largest contribution, 
$ \mathcal{T}(t; \omega_{21}, \omega_{23}, p) \sim \mathcal{O}(\lambda^{-6} T^{-3})$. 
All other contributions are suppressed by a factor of order $\mathcal{O} (\lambda^4)$. 
Note that we later use $\lambda^2 = 10^{-8}$, which implies 
$\mathcal{O} (\lambda^4) \sim 10^{-16}$. This conclusion is corroborated by an 
independent consideration of the large time limit, which we give in appendix \ref{sec:app_c}.
An exemplary plot of the function $\mathcal{T}(t;\omega_{21},\omega_{23},p)$ 
illustrates our findings numerically in figure \ref{fig:data1}.

\subsection{Approximate integration of the lepton number}
\label{sec:int_appr}

\begin{figure}[t]
\centerline{
\includegraphics[width=0.5\textwidth]{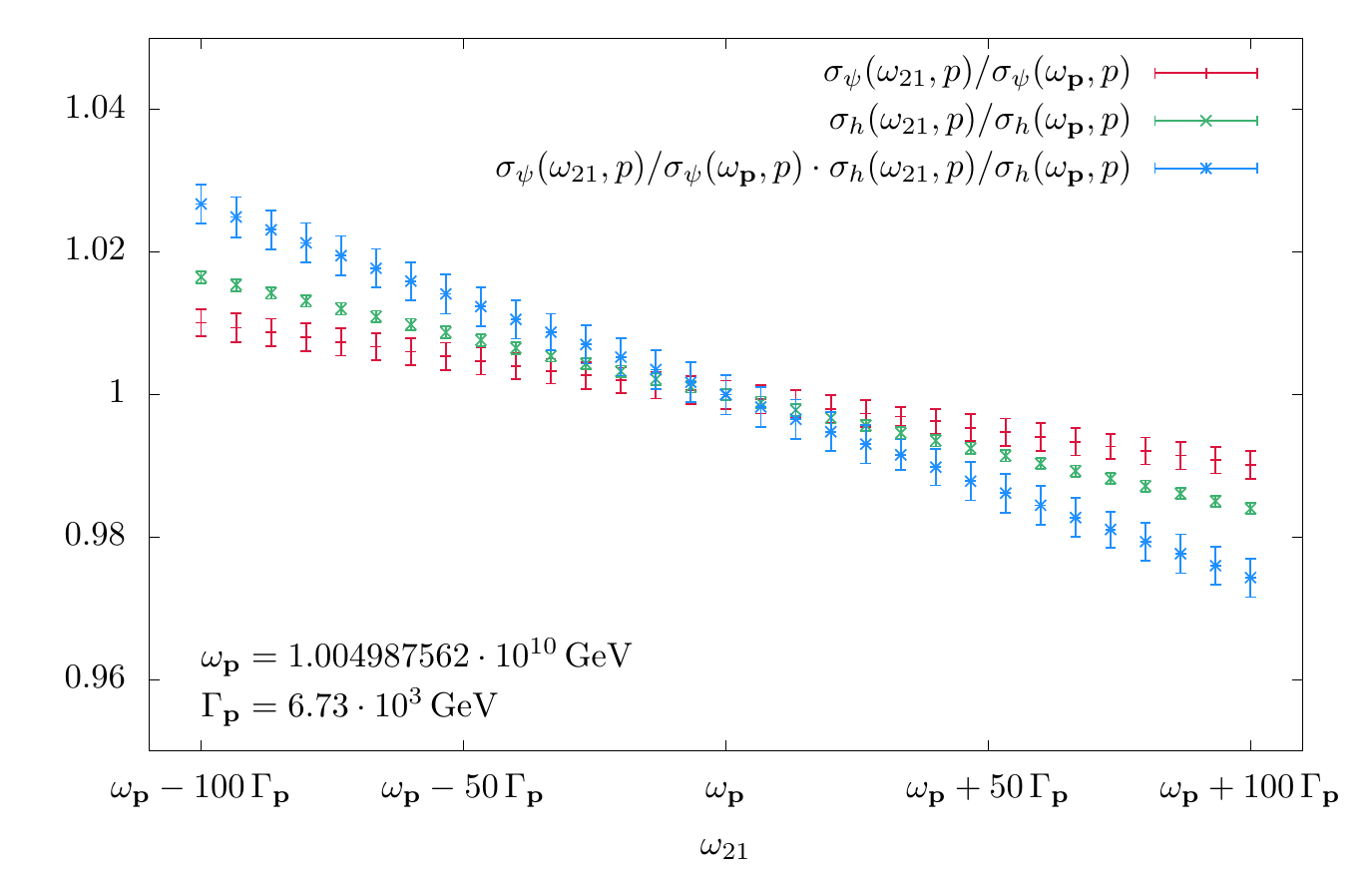}
\includegraphics[width=0.5\textwidth]{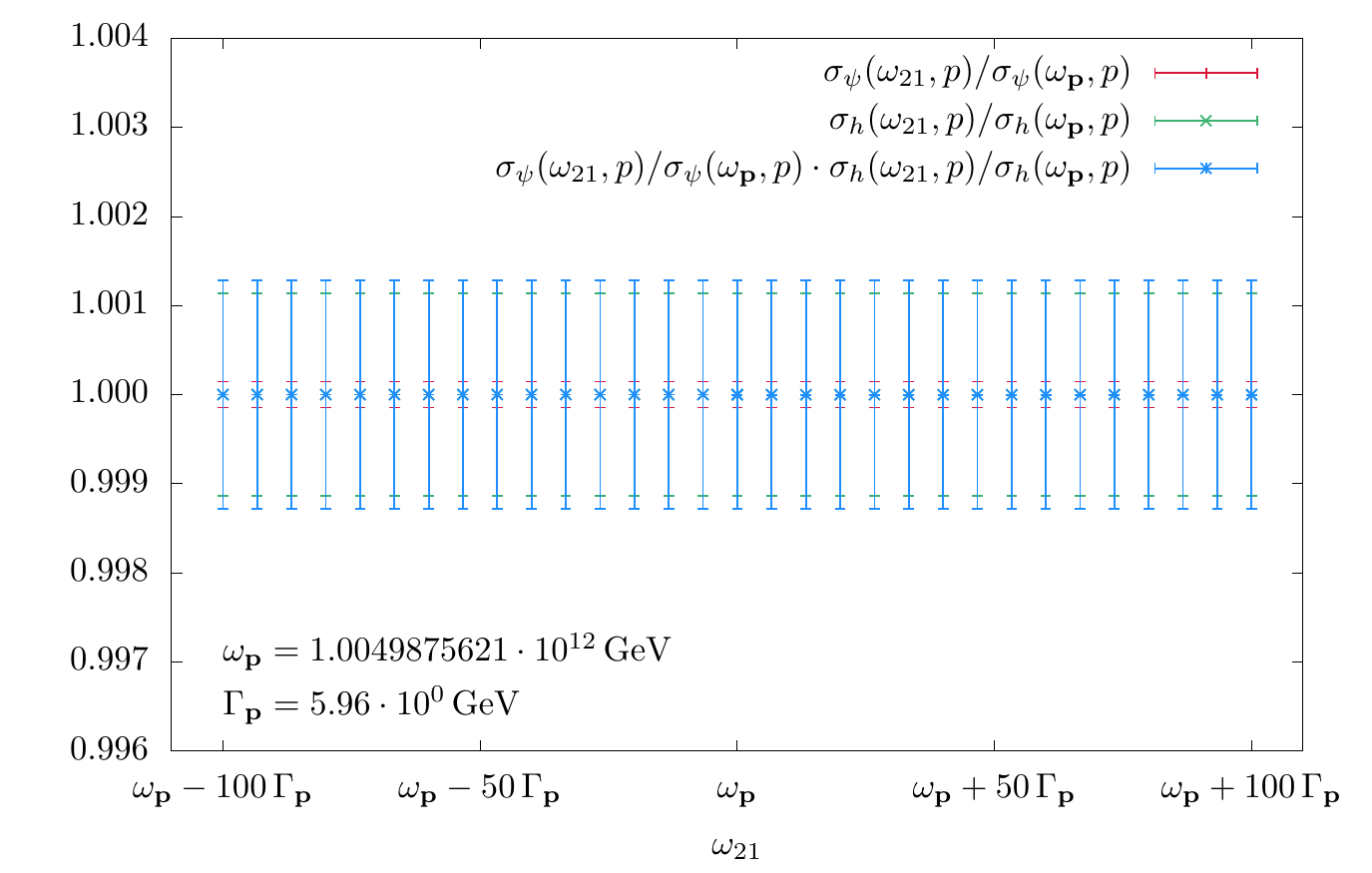}
}
\caption[]{Deviation of the $\sigma_{h,\psi}(\omega_{21},p)$ from the value at $\omega_{21} = \omega_{\bp}$, plotted for $T = 10^{11} \, \mathrm{GeV}$, $M = 10^{10} \, \mathrm{GeV}$ and different $\omega_{\bp}$. \label{fig:data2}}
\end{figure}
We are now going back to the full expression for the lepton number, eq.~\eqref{eq:final}, and factor out the frequency integrals, 
\bea
n_{L,ii}(t) &=& -\frac{128}{\pi} \epsilon_{ii} \int\limits_0^t \diff t_1 \int\limits_0^t \diff t_2 \int\limits_0^{t_2} \diff t_3 \int\limits_{0}^{\infty} \diff p \;
\frac{p^2}{\omega_{\bp}} f_F(\omega_{\bp})\cos(\omega_{\bp} \Delta t_{13}) \text{e}^{-\Gamma_{\bp} \frac{t_{13}}{2}} \nn\\
	&&\times \mathcal{W}(t_1,t_2,t_3;p)\nn\\
\mathcal{W}(t_1,t_2,t_3;p)&=&	
\int\limits_{-\infty}^{\infty} \frac{\diff \omega_{21}}{2 \pi} \int\limits_{-\infty}^{\infty} \frac{\diff \omega_{23}}{2 \pi} f_F (\omega_{21}) \text{Re} \left( \text{e}^{-\text{i}(\omega_{21}\Delta t_{21} + \omega_{23}\Delta t_{23})} \right) \\ \notag &&\times [\sigma_h(\omega_{21}, p)\sigma_h(\omega_{23},p) + \sigma_{\psi}(\omega_{21},p) \sigma_{\psi}(\omega_{23},p)] \;.
\label{eq:w_factor}
\eea
Following the analysis in the previous section, we restrict the frequency integrations to a diagonal strip including the on-shell peak, 
$\omega_{23} \in [-\omega_{21} - a, -\omega_{21} + a]$. We choose $a=\mathrm{const.} \times \Gamma_{\bp}$, such 
that $\omega_{\bp} \gg a \gg \Gamma_{\bp}$ 
 (c.f.\ figure~\ref{fig:data1}). Using the symmetry properties \eqref{eq:sigmaSymm}, we rewrite the $\omega_{21}$ integration to obtain 
\begin{align}
\mathcal{W}(t_1,t_2,t_3;p) 
\label{eq:Weq} \simeq \int\limits_0^{\infty} \frac{\diff \omega_{21}}{2 \pi} &\int\limits_{-\omega_{21}-a}^{-\omega_{21}+a} \frac{\diff \omega_{23}}{2 \pi} \cos(\omega_{21} \Delta t_{21} + \omega_{23} \Delta t_{23}) \\ \notag &\times [\sigma_h(\omega_{21}, p)\sigma_h(\omega_{23},p) + \sigma_{\psi}(\omega_{21},p) \sigma_{\psi}(\omega_{23},p)]\;.
\end{align} 
Next, the $\sigma_{h,\psi}$-functions can be checked numerically to vary at most at the percent level in the interval 
$\omega_{21}, - \omega_{23} \in [\omega_{\bp} - a, \omega_{\bp} + a]$, as depicted in figure~\ref{fig:data2}, with the conservative choice $a = 100 \, \Gamma_{\bp}$.  
The variation is largest for small momenta $p$, which however do not contribute significantly to the integral. 
For the relevant momenta $p \sim \mathcal{O}(T)$, deviations from the on-shell values  
$\sigma_{h,\psi}(\omega_{21} = \omega_{\bp})$ are negligible.
In a second step we may then only keep the leading term in a Taylor expansion around $\omega_{21}=-\omega_{23}=\omega_\bp$ (c.f. eq.~\eqref{eq:sigmaSymm}),
\begin{align}
\sigma_h(\omega_{21},p) &\simeq \sigma_h(\omega_{\bp},p)\;,       &\sigma_{\psi}(\omega_{21}, p) \simeq \sigma_{\psi}(\omega_{\bp},p)\;,\nn\\
\sigma_h(\omega_{23},p) &\simeq \sigma_h(-\omega_{\bp},p) = \sigma_{\psi}(\omega_{\bp},p)\;, &\sigma_{\psi}(\omega_{23}, p) \simeq \sigma_{\psi}(-\omega_{\bp},p) = \sigma_h(\omega_{\bp},p)\;,
\end{align}
allowing us to pull the $\sigma_{h,\psi}$ out of the integration. Note that it is sufficient to consider $+\omega_{\mathbf{p}}$ in this step, because we made use of the symmetry in $\omega_{21}$ before.
We can now do the integration in eq.~\eqref{eq:Weq} analytically leading to
\begin{align}
\mathcal{W} (t_1,t_2,t_3;p)\simeq \frac{\sin(a\Delta t_{23})}{\pi \Delta t_{23}} \delta(\Delta t_{31}) [\sigma_h(\omega_{\bp}, p)\sigma_{\psi}(\omega_{\bp},p)]\;.
\end{align} 
Using this on-shell approximated intermediate result, we are able to carry out the time integration of the lepton number density
\begin{align}
n_{L,ii}(t) \simeq& -\frac{128}{\pi} \epsilon_{ii} \int\limits_0^t \diff t_1 \int\limits_0^t \diff t_2 \int\limits_0^{t_2} \diff t_3 \int\limits_{0}^{\infty} \diff p \;
\frac{p^2}{\omega_{\bp}} f_F(\omega_{\bp}) [\sigma_h(\omega_{\bp}, p)\sigma_{\psi}(\omega_{\bp},p)] \\ \nn &\times \cos(\omega_{\bp} \Delta t_{13}) 
	\text{e}^{-\Gamma_{\bp} \frac{t_{13}}{2}} \frac{\sin(a\Delta t_{23})}{\pi \Delta t_{23}} \delta(\Delta t_{31}) \\ \nn
=& -\frac{128}{\pi} \epsilon_{ii} \int\limits_{0}^{\infty} \diff p \;
\frac{p^2}{\omega_{\bp}} f_F(\omega_{\bp})[\sigma_h(\omega_{\bp}, p)\sigma_{\psi}(\omega_{\bp},p)] \frac{\text{e}^{-\Gamma_{\bp}t}}{4 \pi \Gamma_{\bp}} \bigg[ -\text{i}\bigg( 2\text{Ei}\Big(t(-\text{i}a + \Gamma_{\bp})\Big) \\ \nn 
&- 2 \text{Ei}\Big(t(\text{i}a + \Gamma_{\bp})\Big) - \ln \Big( \frac{-a - \text{i} \Gamma_{\bp}}{a - \text{i}\Gamma_{\bp}} \Big) + \ln \Big( \frac{-a + \text{i}\Gamma_{\bp}}{a + \text{i} \Gamma_{\bp}} \Big) \bigg) + 4 \text{e}^{\Gamma_{\bp}t} \text{Si}(at) \bigg],
\end{align}
where $\text{Ei}(x)$ denotes the exponential integral function and $\text{Si}(x)$ the sine trigonometric integral function. Our choice 
$a=100 \Gamma_{\bp}$ satisfies $a \gg \Gamma_{\bp}$ and allows us to approximately 
consider $at \to \infty$, 
while keeping $\Gamma_{\bp}t$ fixed, leading to an $a$-independent result,
\begin{align}
\label{eq:appFullLept}
n_{L,ii}(t) \simeq -\frac{64}{ \pi} \epsilon_{ii} \int\limits_0^{\infty} \diff p\; \frac{p^2}{\omega_{\bp}} f_F(\omega_{\bp}) \frac{1 - \text{e}^{-\Gamma_{\bp}t}}{\Gamma_{\bp}} \sigma_h(\omega_{\bp}, p)\sigma_{\psi}(\omega_{\bp},p)\;.
\end{align} 
Note that this expression only holds for $t \gtrsim 1/\Gamma_\bp$, since for earlier times the approximation leading to eq.~\eqref{eq:ToffRegion}
is not valid, i.e.\ the region around the on-shell peaks is less suppressed so that the peaks are less pronounced. These   
memory and off-shell effects could of course be kept by a complete numerical integration without
restriction to the regions around $\omega_{21} = - \omega_{23} = \pm \omega_{\bp}$.
Nevertheless, one makes the remarkable observation that at sufficiently late times
our result has the same time dependence as the solution of the corresponding Boltzmann equations \cite{Anisimov:2010dk}, but
this time based on a full calculation without any assumptions.

Ultimately, we are interested in the value of the generated lepton number once it is thermalised. 
This is easily obtained by taking the limit $t \to \infty$, resulting in
\begin{align}
\label{eq:appThermLep}
n_{L,ii}^T \simeq \lim_{t \to \infty} n_{L,ii} (t,t) = -\frac{64}{ \pi} \epsilon_{ii} \int\limits_0^{\infty} \diff p \;\frac{p^2}{\omega_{\bp}} f_F(\omega_{\bp}) \frac{1}{\Gamma_{\bp}} [\sigma_h(\omega_{\bp}, p)\sigma_{\psi}(\omega_{\bp},p)]\;.
\end{align} 
As a valuable crosscheck for our approximate calculation, we change the order of integration in  appendix \ref{sec:app_c}, i.e.\ we first do
the time integrals, followed by the $t\to \infty$ limit and only then integrate over the frequencies, arriving at the same result.

Finally, it is also interesting to quantify the timescale for thermalisation, which is expected to be of order
$t_T \sim \mathcal{O}(1/\Gamma_{\bp})$ according to the discussion in section~\ref{sec:scenario}. 
We define $t_T$ as the time where
\begin{align}
n_{L,ii} (t_T) = \left( 1 - \text{e}^{-1} \right) n_{L,ii}^T\;,
\end{align}
which gives $t_T = 1/\Gamma$ for a constant $\Gamma_\bp \equiv \Gamma$, cf.\ eq.~\eqref{eq:appFullLept}.

\subsection{Time evolution of the lepton number}

The final integration over the Majorana neutrino momentum is performed numerically using integration and ODE algorithms implemented
in the \textit{GNU scientific library} \cite{contributors-gsl-gnu-2010} and the \textit{BOOST library} \cite{schling2014boost}. 
Following \cite{Anisimov:2010dk}, we consider a Majorana neutrino mass of $M = 10^{10} \, \mathrm{GeV}$ and a constant 
temperature of $T = 10^{11} \, \mathrm{GeV}$ for the time-dependent calculation. This choice is consistent with the physical scenario outlined
in section~\ref{sec:scenario}, allowing to neglect the Hubble expansion.
For the coupling of the corrected Majorana neutrino decay width we choose $\lambda^2 = 10^{-8}$, in analogy 
to \cite{Anisimov:2010gy}.\footnote{In fact our presentation is independent of the choice of 
$\lambda^2$ because it is possible to rescale $t$ in eq.\ \eqref{eq:appFullLept} accordingly and factor $\lambda^2$ out 
of $1/\Gamma_{\bp}$ together with $\lambda_{ii}$.} 
The SM couplings are obtained by solving the renormalization group equations from \cite{Schrempp:1996fb, Arason:1991ic} using the
 SM parameters from \cite{Patrignani:2016xqp}. In analogy to the calculation of the asymptotic masses, we have neglected all contributions
 from the quark sector except for the top quark. All details are given in appendix~\ref{sec:app_RGEs}.
\begin{figure}[t]
\centerline{
\includegraphics[width=0.5\textwidth]{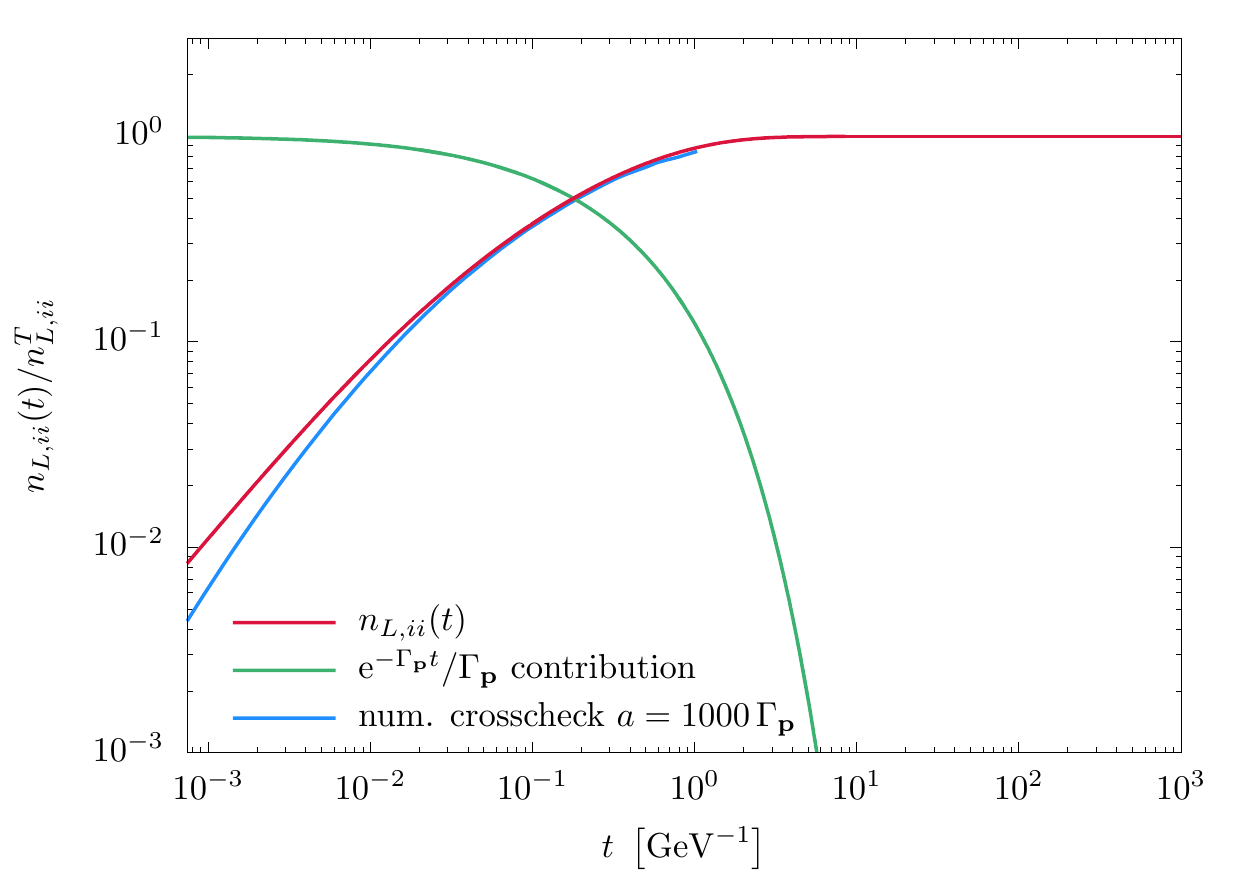}
\includegraphics[width=0.5\textwidth]{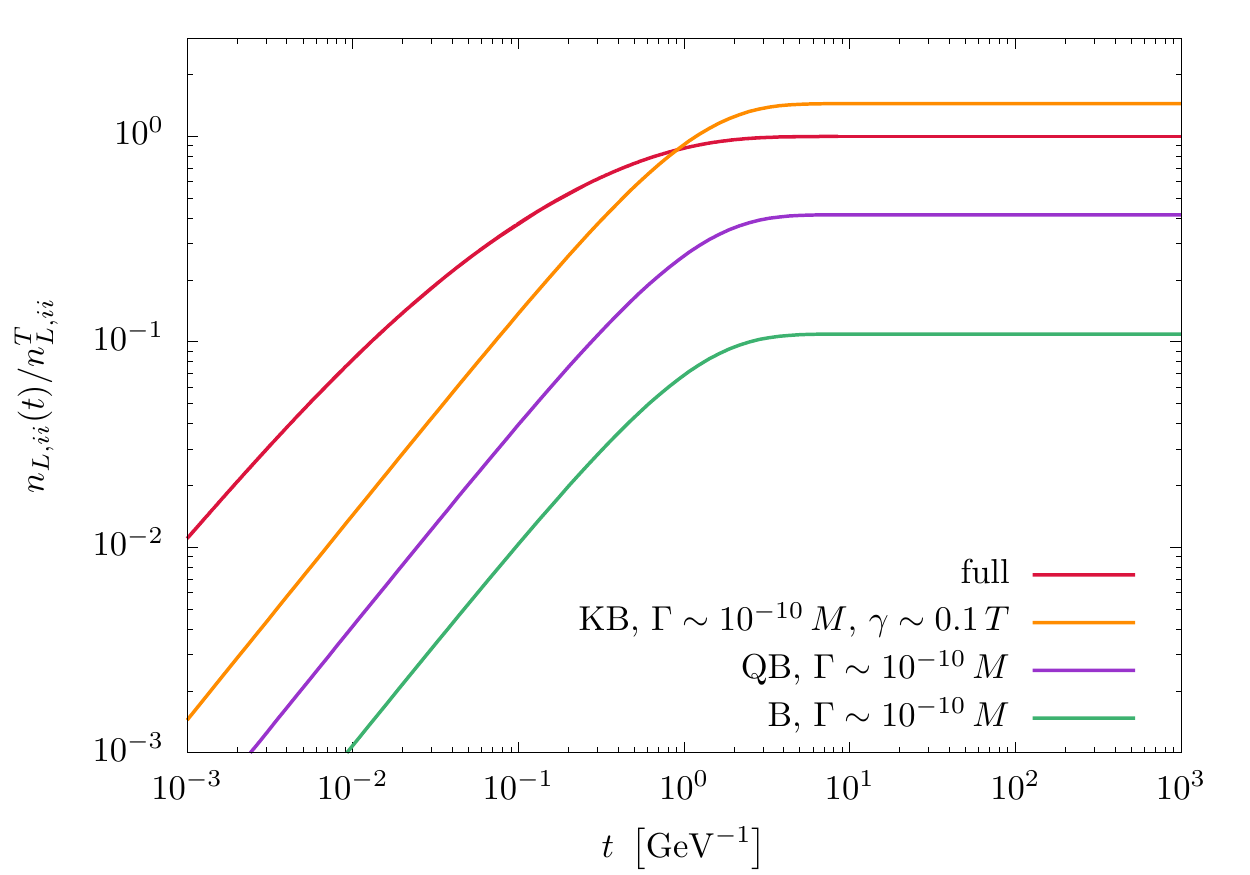}
}
\caption[]{Time evolution of the lepton number density. Left: Full result obtained by different orders of integration
(see text). The green curve gives the time-dependent contribution. Right: Comparison of the full result with Boltzmann (B), quantum Boltzmann (QB) and Kadanoff-Baym (KB) computations, where gauge corrections are not computed but parametrised by thermal widths.\label{fig:t-dep}}
\end{figure}

The resulting time evolution of the lepton number is shown in figure~\ref{fig:t-dep} (left) by the red curve. The blue curve corresponds
to an alternative evaluation of the integrals, where we start from the time-integrated expression, $\mathcal{T}(t;\omega_{21},\omega_{23}, p)$ 
in appendix \ref{sec:app_b}, and numerically integrate over the frequencies with
$\omega_{23} \in [-\omega_{21} - a, -\omega_{21} + a]$, this time with the larger range $a=1000 \, \Gamma_\bp$
and the full functions $\sigma_{h,\psi}$.
As the figure shows, this crosscheck fully confirms the validity of our approximations at sufficiently large times. 
Deviations between the two curves  show up only at early times, when the on-shell peak of the integrand is less pronounced. 
The green curve shows the time-dependent term in eq.~\eqref{eq:appThermLep}, which dies away exponentially, leaving us with the fully thermalised
value of the lepton number. The thermalisation time in the figure is consistent with the expectation for our choice of parameter values,
i.e.~we have $\Gamma_\bp\sim \mathcal{O}(\mathrm{GeV})$ and $t_T\sim  \mathcal{O}(1/\mathrm{GeV})$.

In figure~\ref{fig:t-dep} (right) we compare our complete result to various other approaches discussed previously \cite{Anisimov:2010dk}. 
These are the solutions of Boltzmann equations (B), Boltzmann equations using quantum mechanical distribution functions (QB), and
Kadanoff-Baym equations without gauge corrections, but with thermal widths for the lepton and Higgs propagators, 
$\gamma=\gamma_l+\gamma_\phi$, introduced by hand (KB).\footnote{To be specific, we use eq.~(7.2)  from \cite{Anisimov:2010dk} and 
perform an additional $\mathbf{k}$-integration, inserting
eq.~(7.3) for (B),  eq.~(7.10) for (QB) and eq.~(7.8) for (KB).}
In all three cases we work with a constant decay width for the Majorana neutrino propagator, 
$\Gamma_\bp(\omega_{\mathbf{p}})=\Gamma\sim 10^{-10} M \sim \mathcal{O}(\mathrm{GeV})$, corresponding to the dominant contribution in
 our full calculation.
 For the SM thermal damping widths we choose $\gamma \sim 0.1 ~ T$ as expected from 
thermal field theory \citep{Bellac:2011kqa}. 
All results are normalised by the value of the thermalised lepton number density from the full calculation.
While all curves have the same qualitative features, the Boltzmann result deviates by almost an order of magnitude, the quantum 
Boltzmann computation reduces the difference significantly and 
Kadanoff-Baym with thermal widths is $\mathcal{O}(1)$ accurate. 
The qualitative similarity is due to eq.~\eqref{eq:appFullLept} having the same explicit time dependence as
the Boltzmann solutions. The quantitative difference of the full solution stems from the functions $\sigma_{h,\psi}$ 
and a momentum-dependent $\Gamma_\bp$, effecting different changes for the weights of the momentum modes.
We may thus conclude that gauge corrections indeed cause the 
expected narrow damping widths at late times. 
However, as expected in \cite{Anisimov:2010dk,Anisimov:2010aq}, for quantitative results the full calculation is mandatory.

\subsection{Temperature dependence of the gauge corrections}

\begin{figure}[t]
\centerline{
\includegraphics[width=0.5\textwidth]{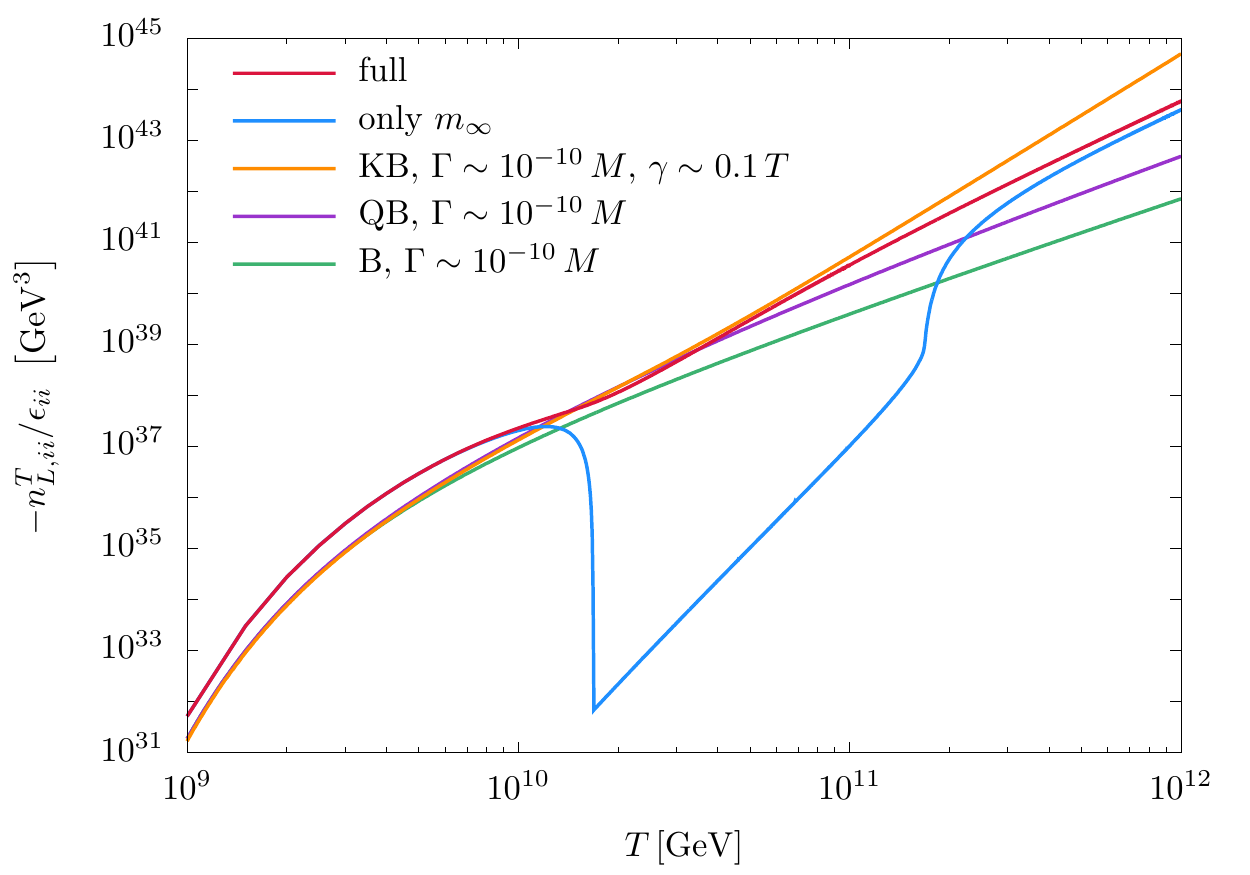}
\includegraphics[width=0.5\textwidth]{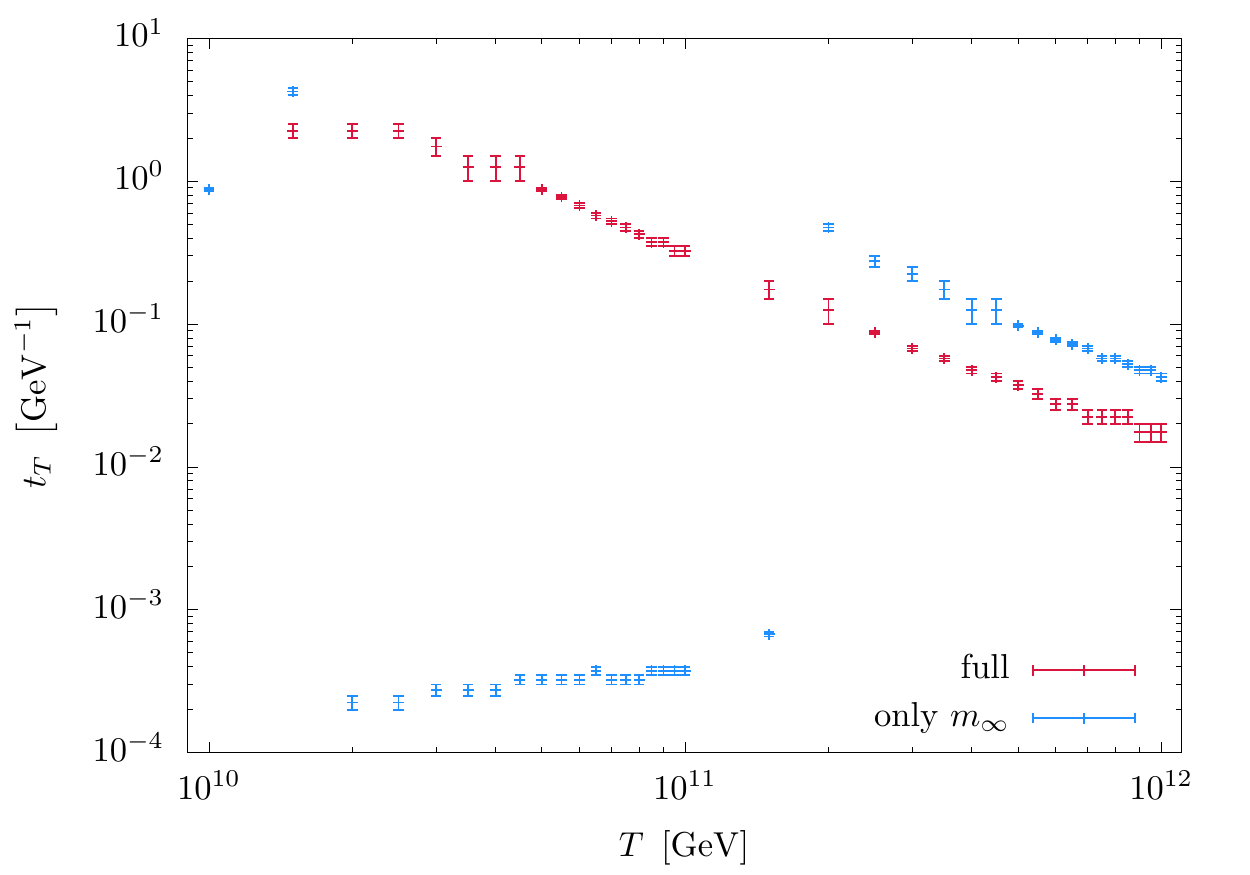}
}
\caption[]{Left: Thermalised lepton number density from the full resummation, only including asymptotic masses, Boltzmann (B), 
quantum Boltzmann (QB) and Kadanoff-Baym with parametrised propagators (KB), as a function of temperature.
Right: Thermalisation time as a function of temperature. \label{fig:T}}
\end{figure}
In the simplest leptogenesis scenario considered here, the temperature is constant
to first approximation. However, gauge corrections apply to other parameter choices and scenarios as well, and it is interesting to investigate
their relative importance as a function of temperature. In figure~\ref{fig:T} (left) we repeat the comparison of the previous section in case of the thermalised lepton number as a function of temperature $T$. Only including asymptotic masses leads to a kinematical suppression
of the Majorana decay and production rate \cite{Anisimov:2010gy}, and hence of the lepton number, in some temperature range,
which is unphysical. This illustrates 
the importance of a complete calculation including the resummed gauge corrections. At yet higher temperatures, 
once $T \gg M$, the kinematic suppression is switched off and only including asymptotic masses gives a good 
approximation of the full result.

For low temperatures $T\lsi M$, the results of Boltzmann, quantum Boltzmann and Kadanoff-Baym with 
thermal damping widths are very close. This agrees with the smallness of corrections observed in the non-relativistic 
case \citep{Bodeker:2017deo}.
Since the resummed gauge corrections are suppressed in this regime, our calculation is close to the previous results as well.
Note however, that the agreement cannot be exact,
because the weights of the momentum modes are still not identical,
due to the momentum dependent $\Gamma_\bp$. 
It is thus the choice of the (momentum independent) model parameters $\Gamma$ and $\gamma$ 
which determines the temperature range of best 
agreement between the gauge resummed and approximate solutions. For the choice presented here this is at $T\sim M$. 
Another choice achieves full agreement at low temperatures at the expense of larger discrepancies at high temperatures,
while an appropriately temperature dependent $\Gamma$ can increase the range of agreement.

For temperatures $T>M$, the different solutions split up. 
Introducing thermal damping widths by hand repairs the qualitative deficit of kinematic suppression (and agrees well with the full result in a part 
of this region), but overestimates the lepton number for high temperatures. This is because 
the dominant momenta become more light-like, and hence the resummation more important. 
Note however that, by altering the scaling of the Majorana neutrino decay width to  e.g.~$\Gamma \sim 10^{-11} \, T$,
better agreement with the full result is achieved.

Finally, figure~\ref{fig:T} (right) shows the thermalisation times for the lepton number, extracted from the full result and the one including 
asymptotic masses only.  
Apart from the region with kinematic suppression we roughly find a scaling $t_T \sim \mathcal{O} (1/\Gamma_\bp) \propto 1/T$ as expected.

\section{Conclusions}
 
In this work the leading-order gauge corrections to quantum leptogenesis were included in a fully quantum field theoretical calculation 
based on non-equilibrium Kadanoff-Baym equations. In this setting, gauge corrections appear first at the three-loop level 
of lepton self-energies. In the symmetric electroweak phase, gauge bosons are massless on tree-level and, 
for $T \gtrsim M$ and collinear thermal loop momenta, 
require resummations to infinite loop order to complete a leading-order calculation in the gauge coupling.
We have adapted a previous CTL resummation of the Majorana neutrino self-energy in such a way that it can be applied to the 
lepton self-energy graphs relevant for quantum leptogenesis. Our resulting lepton number density, eq.~\eqref{eq:final}, is complete to 
leading order in all SM couplings  and forms the basis for a quantitatively accurate evaluation of leptogenesis. 

We have evaluated our expression for the simplest vanilla leptogenesis scenario at late times approaching thermalisation, 
which allows for an analytic analysis and integration of
the lepton number density in Fourier space.
One observes how the gauge corrections dynamically cause a dominant peak to grow in the integrand, whose width at late times 
corresponds to the thermal decay width of the Majorana neutrino. Similarly, thermal widths for leptons and Higgs propagators are 
automatically and fully included. The resulting late time behaviour qualitatively agrees with that observed from Boltzmann 
equations.
However, for quantitative results, the full quantum calculation is necessary.
Finally, we have evaluated the temperature dependence
of the gauge corrections and the thermalisation time to estimate effects on other parameter choices or scenarios.  
Boltzmann equations dressed with quantum distribution 
functions underestimate the full result by $\mathcal{O}(1)$ for $T \gtrsim M$, introducing an uncertainty of similar size as that of spectator processes. 
Thus, we have also obtained a first complete estimate of the theoretical uncertainties for leptogenesis.
On the other hand, for $T\lesssim M$ the resummed gauge corrections are suppressed and our result is close
to the previous ones.

Our  calculation can be easily adopted to other parameter regions or modfied scenarios, such as resonant leptogenesis.
For yet better accuracy, the calculation could be further refined by also including washout diagrams as well as the 
temperature change due to the Hubble expansion.

\acknowledgments

We thank D.~B\"odeker and W.~Buchm\"uller for useful discussions and suggestions.
Numerical calculations were performed on the LOEWE-CSC computer at Goethe University Frankfurt.
This work is supported by the ERC Starting Grant `NewAve' (638528) as well as by the 
Deutsche Forschungsgemeinschaft under Germany's Excellence Strategy -- EXC 2121 `Quantum Universe' -- 390833306.

\newpage
\appendix
\section{Feynman diagrams for gauge corrections at three-loop level}
\label{sec:app_a}

\centerline{
\includegraphics[width=\textwidth]{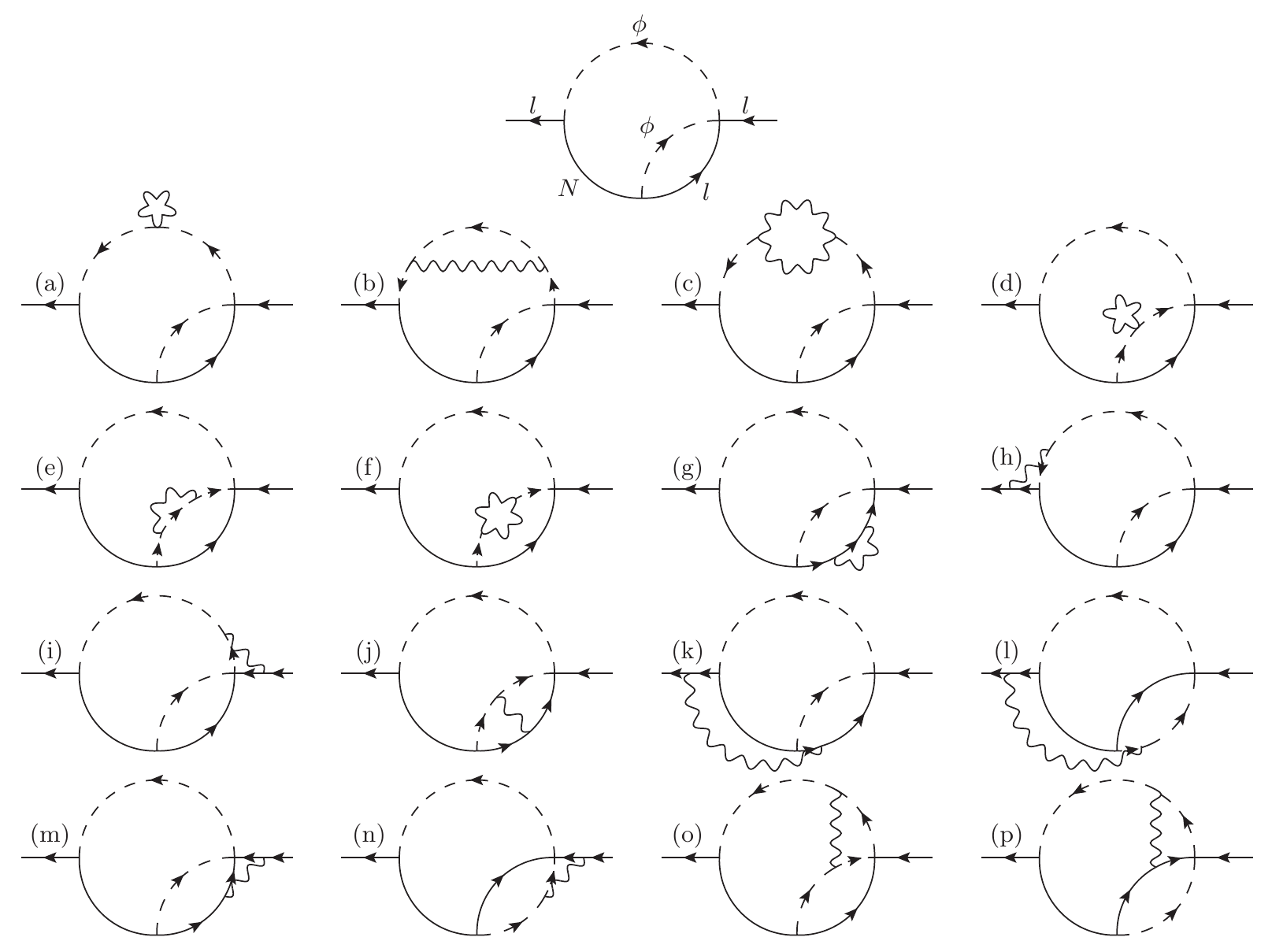}
}
Here we list all Feynman diagrams obtained by adding one gauge boson line to the diagrams in figure~\ref{fig:CpViolation}.
The diagrams (a)-(g) are gauge corrections to the lepton and Higgs propagators, 
which are included by the asymptotic masses, eq.~\eqref{eq:m_as}.
The diagrams (h)-(j) are vertex gauge corrections, which are 
included in the resummed ladder diagram figure \ref{fig:convert_diag}.
The remaining diagrams 
are further suppressed because the mean free path between heavy neutrino vertices exceeds the Debye screening length,
for details see appendix \ref{sec:app_resum}.

\section{Leading and sub-leading gauge corrections}
\label{sec:app_resum}

\begin{figure}[H]
\centerline{
\includegraphics[width=\textwidth]{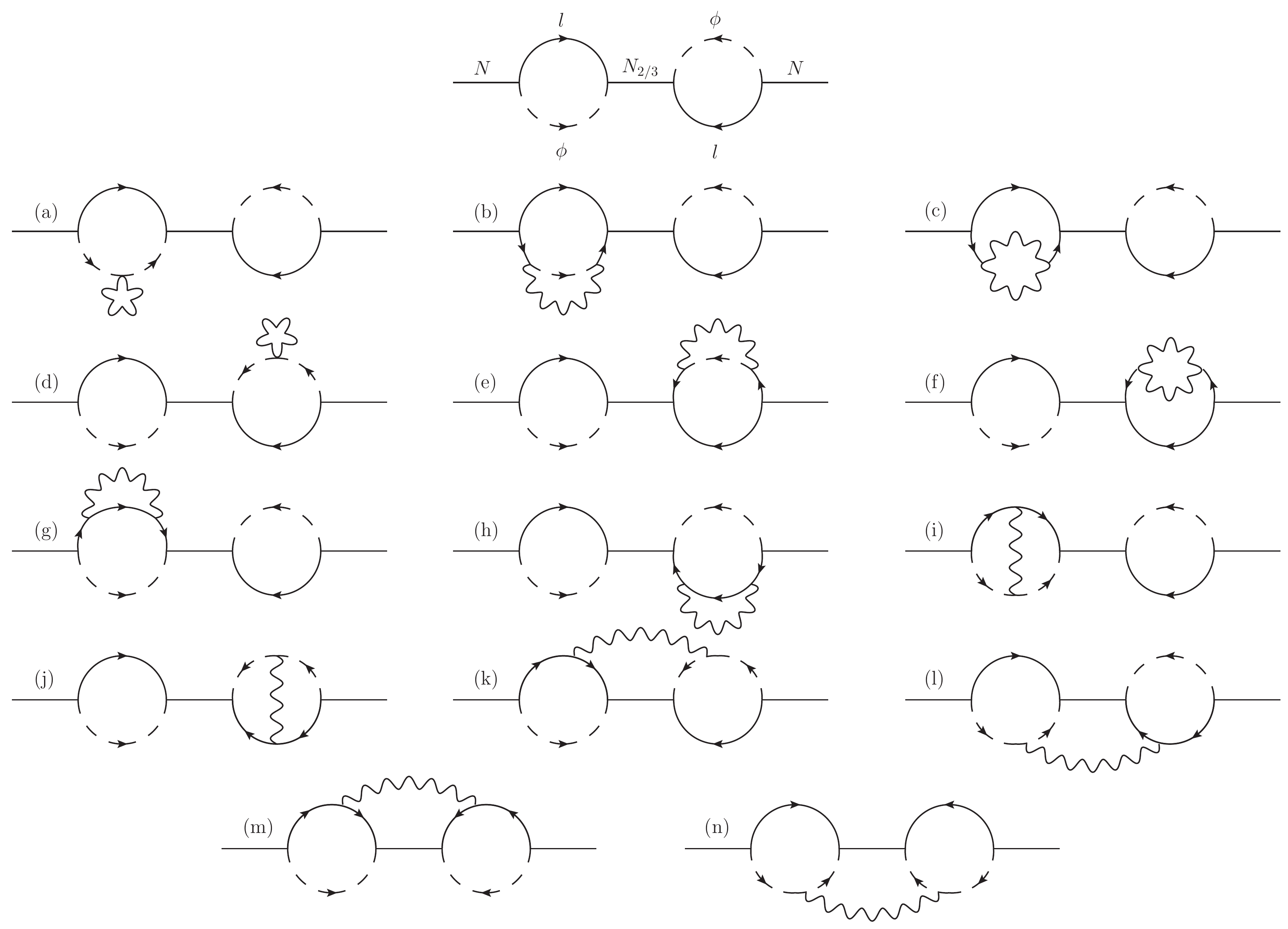}
}
\vspace{-0.3cm}
\caption[]{Gauge corrections to the propagator-type self-energy contribution of $N$.} 
\label{fig:wave}
\vspace{0.5cm}
\centerline{
\includegraphics[width=0.75\textwidth]{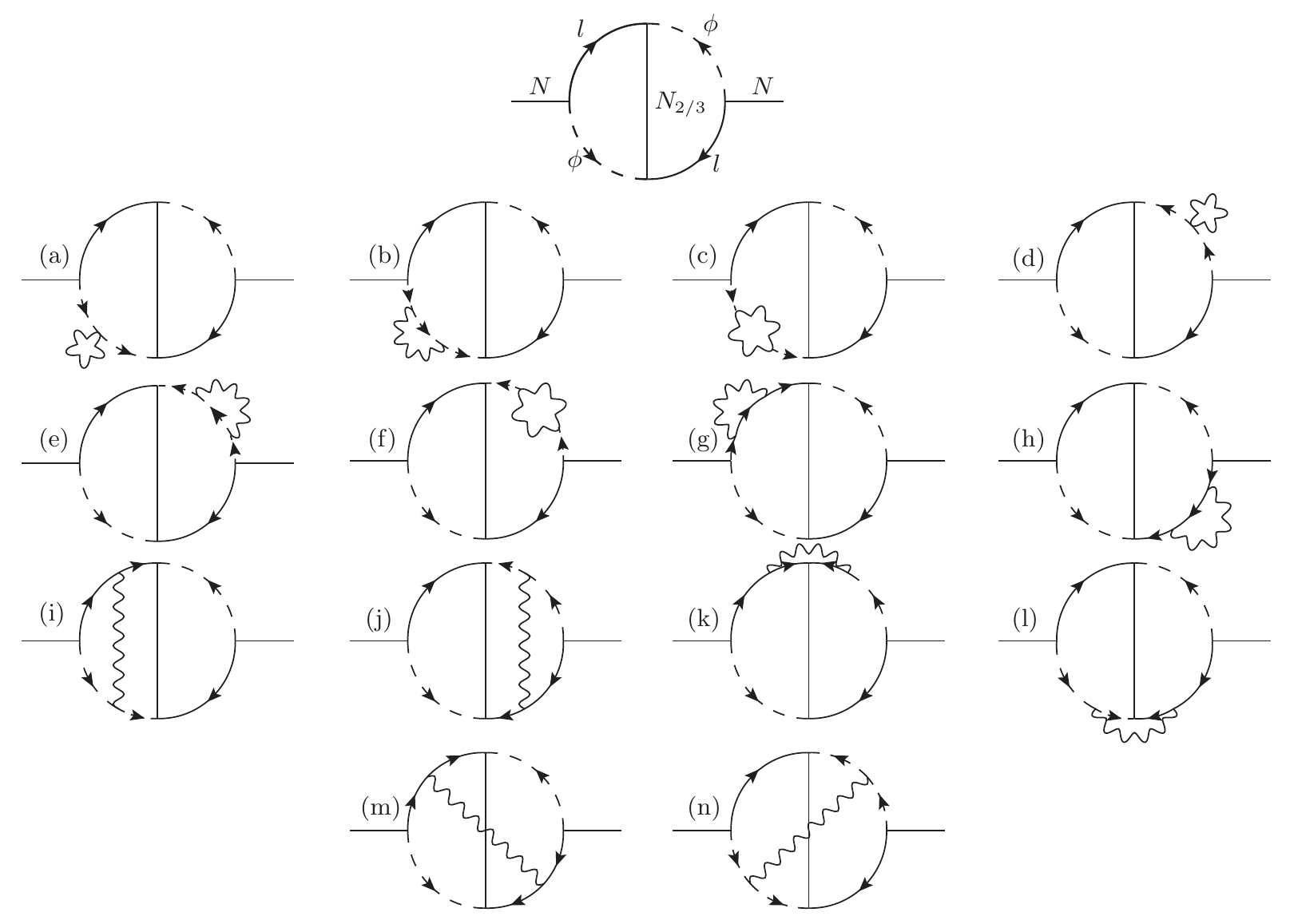}
}
\vspace{-0.3cm}
\caption[]{Gauge corrections to the vertex-type self-energy contribution of $N$.}
\label{fig:vertex}
\end{figure}

In order to distinguish leading and sub-leading gauge corrections, as well as for generalisations to other neutrino mass configurations,
it is useful to step back from the hierarchical scenario and re-introduce the heavy neutrinos $N_{2,3}$. 
The self-energy diagrams of the  neutrino $N$ can then be separated into two classes, which can be described as
either "propagator-type" or "vertex-type" corrections, figures \ref{fig:wave} and \ref{fig:vertex}, respectively.

A parametric suppression applies to a diagram whenever the mean free path between emission and absorbtion of a soft gauge
boson exceeds its associated screening length \cite{Arnold:2002ja}. In our case, the mean free path of the heavy neutrinos is 
$l_{N_i}\sim (\lambda^2M_i)^{-1}$, while the Debye screening length is $l_D\sim (gT)^{-1}$. We thus have $l_N\gg l_D$ as long as
\beq
M_i\ll \frac{g}{\lambda^2}T\;.
\eeq
\newpage

With $\lambda^2\sim 10^{-8}$, this is satisfied for all mass scenarios provided the $M_{i}$ are sufficiently below
the Planck scale. Hence, the diagrams (k)-(n) in figure \ref{fig:wave} and (i), (j), (m), (n) in figure \ref{fig:vertex} are parametrically suppressed.
A similar suppression occurs for vertex corrections as (k), (l) in figure \ref{fig:vertex}, once the $N$-line is closed. All remaining
diagrams have to be resummed, as explained in section \ref{sec:CTLResummation}. Finally, we return to the hierarchical scenario
and integrate out the $N_{2,3}$, where both propagator-type and vertex-type diagrams reduce to a double blob diagram with 
an effective four-point vertex.
Upon closing the external lines we then obtain the resummed diagram in figure \ref{fig:convert_diag}.

\section{The function $\mathcal{T}(t;\omega_{21},\omega_{23},p)$}
\label{sec:app_b}

The integral in eq.~\eqref{eq:timeDepPart} can be performed analytically giving the result
\begin{align}
\label{eq:Tfactor}
&\mathcal{T}(t; \omega_{21}, \omega_{23}, p) = \frac{\text{e}^{-\Gamma_\bp t}}{(\omega_{21} + \omega_{23}) (\Gamma_\bp^2 + 4 (\omega_{23} - \omega_{\bp})^2) (\Gamma_\bp^2 + 4 (\omega_{23} + \omega_{\bp})^2)} \notag \\
&\times \frac{1}{(\Gamma_\bp^2 + 4 (\omega_{21} - \omega_{\bp})^2)  (\Gamma_\bp^2 + 4 (\omega_{21} + \omega_{\bp})^2)} \bigg\{8 \Gamma_\bp (\omega_{21} + \omega_{23}) \notag \\
&\times \Big[-(\Gamma_\bp^2 +
4 \omega_{21}^2) (\Gamma_\bp^2 + 4 \omega_{23}^2) -
8 (\Gamma_\bp^2 +
2 (\omega_{21}^2 -
4 \omega_{21} \omega_{23} + \omega_{23}^2)) \omega_{\bp}^2 -
16 \omega_{\bp}^4\Big] \notag \\
&- 8 \text{e}^{\Gamma_\bp t} \Gamma_\bp (\omega_{21} +
\omega_{23}) \Big[(\Gamma_\bp^2 +
4 \omega_{21}^2) (\Gamma_\bp^2 + 4 \omega_{23}^2) +
8 (\Gamma_\bp^2 +
2 (\omega_{21}^2 -
4 \omega_{21} \omega_{23} + \omega_{23}^2)) \omega_{\bp}^2 +
16 \omega_{\bp}^4\Big] \notag \\
&\times \cos(t (\omega_{21} + \omega_{23})) +
4 \text{e}^{\Gamma_\bp t} (\Gamma_\bp^2 -
4 \omega_{21} \omega_{23} +
4 \omega_{\bp}^2) \sin (t (\omega_{21} + \omega_{23})) \notag \\
&\times \Big[\Gamma_\bp^4 +
16 (\omega_{21} - \omega_{\bp}) (\omega_{23} - \omega_{\bp})
(\omega_{21} + \omega_{\bp}) (\omega_{23} + \omega_{\bp}) +
4 \Gamma_\bp^2 (\omega_{21}^2 + \omega_{23}^2 +
2 \omega_{\bp}^2)\Big] \notag \\
&+ 8 \text{e}^{\Gamma_\bp t/2} \Big[\cos(t (\omega_{21} - \omega_{23})/2) \cos(t \omega_{\bp})
 \Big(2 \Gamma_\bp (\omega_{21} + \omega_{23})
((\Gamma_\bp^2 + 4 \omega_{21}^2) (\Gamma_\bp^2 +
4 \omega_{23}^2) \notag \\
&+ 8 (\Gamma_\bp^2 +
2 (\omega_{21}^2 -
4 \omega_{21} \omega_{23} + \omega_{23}^2))
\omega_{\bp}^2 + 16 \omega_{\bp}^4) \cos(
t (\omega_{21} + \omega_{23})/2) \notag \\
&- (\Gamma_\bp^4 +
16 (\omega_{21} - \omega_{\bp}) (\omega_{23} - \omega_{\bp}) (
\omega_{21} + \omega_{\bp}) (\omega_{23} + \omega_{\bp}) +
4 \Gamma_\bp^2 (\omega_{21}^2 + \omega_{23}^2 +
2 \omega_{\bp}^2)) \notag \\
&\times (\Gamma_\bp^2 -
4 \omega_{21} \omega_{23} +
4 \omega_{\bp}^2) \sin(
1/2 t (\omega_{21} + \omega_{23}))\Big) -
4 (\omega_{21} - \omega_{23}) \omega_{\bp} \sin(t (\omega_{21} - \omega_{23})/2) \notag \\
&\times \Big(-4 \Gamma_\bp (
\omega_{21} + \omega_{23}) (\Gamma_\bp^2 -
4 \omega_{21} \omega_{23} + 4 \omega_{\bp}^2) \cos(
t (\omega_{21} + \omega_{23})/2) \notag \\
&+ (\Gamma_\bp^4 +
16 (\omega_{21} - \omega_{\bp}) (\omega_{23} - \omega_{\bp}) (
\omega_{21} + \omega_{\bp}) (\omega_{23} + \omega_{\bp}) -
4 \Gamma_\bp^2 (\omega_{21}^2 +
4 \omega_{21} \omega_{23} + \omega_{23}^2 -
2 \omega_{\bp}^2)) \notag \\
&\times \sin(
1/2 t (\omega_{21} + \omega_{23}))\Big) \sin(
t \omega_{\bp})\Big]\bigg\}\;.
\end{align}

\section{Late-time limit for the lepton number}
\label{sec:app_c}
Here we provide a crosscheck for the derivation of the approximated lepton number 
density in the infinite-time limit.
Starting from the full expression, eq.~\eqref{eq:t-int},
taking the infinite time limit in the time dependent part $\mathcal{T}(t; \omega_{21}, \omega_{23}, p)$ of the integrand reads
\begin{align}
\lim_{t \to \infty}  \mathcal{T}(t; \omega_{21}, \omega_{23}, p) = \frac{4 \pi (\Gamma_\bp^2 + 4(\omega_{\bp}^2 + \omega_{21}^2))}{\Gamma_\bp^4 + 16(\omega_{\bp}^2 - \omega_{21}^2)^2 + 8 \Gamma_\bp^2(\omega_{\bp}^2 + \omega_{21}^2)} \delta(\omega_{21} + \omega_{23})\;.
\end{align}
Note the explicit appearance of a Delta function enforcing $\omega_{23} = -\omega_{21}$, which is consistent 
with the dominant on-shell approximation $\omega_{21} = \pm \omega_{\bp}$, 
$\omega_{23} = \mp \omega_{\bp}$ identified in section \ref{sec:t_int}.

Carrying out the $\omega_{23}$ integration in the infinite-time limit and using the 
symmetry properties of the $\sigma$-functions~\eqref{eq:sigmaSymm}, we arrive at
\begin{align}
\lim_{t \to \infty} n_{L,ii}(t) = -\frac{128}{\pi^2} \epsilon_{ii} \int\limits_0^{\infty} dp \int\limits_{-\infty}^{\infty} \frac{d \omega_{21}}{2 \pi} &\frac{p^2}{\omega_{\bp}} \frac{4 \pi f_F(\omega_{\bp}) (\Gamma_\bp^2 + 4(\omega_{\bp}^2 + \omega_{21}^2))}{\Gamma_\bp^4 + 16(\omega_{\bp}^2 - \omega_{21}^2)^2 + 8 \Gamma_\bp^2(\omega_{\bp}^2 + \omega_{21}^2)} \\ \notag
&\times \sigma_{\psi}(\omega_{21}, p) \sigma_h(\omega_{21}, p)\;.
\end{align}
Approximating the $\sigma_{h,\psi}$-functions by their on-shell contributions, cf.\ section~\ref{sec:int_appr},
\begin{align}
\sigma_{\psi}(\omega_{21}, p) \approx \sigma_{\psi}(\omega_{\bp}, p)\;,~~~~\sigma_h(\omega_{21}, p) \approx \sigma_h(\omega_{\bp}, p)\;,
\end{align}
we can do the $\omega_{21}$-integral to recover
\begin{align}
\lim_{t \to \infty} n_{L,ii}(t) = -\frac{64}{\pi} \epsilon_{ii} \int\limits_0^{\infty} dp \frac{p^2}{\omega_{\bp}} f_F(\omega_{\bp}) \frac{1}{\Gamma_\bp} \sigma_{\psi}(\omega_{\bp}, p) \sigma_h(\omega_{\bp}, p)\;,
\end{align}
as in eq.~\eqref{eq:appThermLep}, validating our approximate integration in section~\ref{sec:int_appr}.

\section{SM parameters for the numerical evaluation}
\label{sec:app_RGEs}

In order to evolve the SM parameters to the scale given by the temperature $\mu = 2 \pi T$, we employ the renormalization group equations (RGEs) from~\cite{Schrempp:1996fb, Arason:1991ic} taking only into account the largest coupling to a quark, i.e.\ the top quark being the heaviest quark,
\begin{align}
\frac{\diff g^{\prime 2}}{\diff \tau} &= \frac{g^{\prime 4}}{8 \pi^2} \frac{41}{10} + \mathcal{O}(g_{\text{SM}}^6)\;, \\
\frac{\diff g^2}{\diff \tau} &= \frac{g^4}{8 \pi^2} \left( - \frac{19}{6} \right) + \mathcal{O}(g_{\text{SM}}^6)\;, \\
\frac{\diff g_s^2}{\diff \tau} &= \frac{g_s^4}{8 \pi^2} (-7) + \mathcal{O}(g_{\text{SM}}^6)\;, \\
\frac{\diff h_t^2}{\diff \tau} &= \frac{h_t^2}{8 \pi^2} \left( \frac{9}{2} h_t^2 - \frac{17}{20} g^{\prime 2} - \frac{9}{4} g^2 - 8 g_s^2 \right) + \mathcal{O}(g_{\text{SM}}^6)\;, \\
\frac{\diff \lambda_\phi}{\diff \tau} &= \frac{1}{16 \pi^2} \left( \frac{27}{200} g^{\prime 4} + \frac{9}{20} g^{\prime 2} g^2 + \frac{9}{8} g^4 - \frac{9}{5} g^{\prime 2} \lambda_\phi - 9 g^2 \lambda_\phi \right. \nonumber \\
&\left. \ \ \ \ \ \ \ \ \ \ \ \ \ - 6 h_t^4 + 12 h_t^2 \lambda_\phi + 24 \lambda_\phi^2 \right) + \mathcal{O} (g_{\text{SM}}^6)\;,
\end{align}
where $\tau \equiv \ln (\mu / \mu_0)$ and we choose $\mu_0 = 2 \pi T_0$ with $T_0 = 10^9 \, \mathrm{GeV}$. For notational simplicity $\mathcal{O} (g_{\text{SM}}^6)$ denotes any combination of SM couplings and $\lambda_\phi \sim \mathcal{O} (g_{\text{SM}}^2)$ for this counting. As initial conditions we use at $\tau_Z \equiv \ln (m_Z/\mu_0)$~\cite{Patrignani:2016xqp}
\begin{align}
m_Z &=91.1876(21) \, \mathrm{GeV}\;, \\
\alpha_{em} (\tau_Z) &= 1/127.950(17)\;, \\
\alpha_s (\tau_Z) &= 0.1182(16)\;, \\
\sin^2 \theta_W (\tau_Z) &= 0.23129(5)\;, \\
m_t &= 173.21(1.22) \, \mathrm{GeV}\;, \\
m_H &= 125.09(24)  \, \mathrm{GeV}\;, \\
G_F &= 1.1663787(6) \cdot 10^{-5} \, \mathrm{GeV}^{-2} \Rightarrow v = \tfrac{1}{\sqrt{\sqrt{2} G_F}} = 246.2197(1) \, \mathrm{GeV}\;.
\end{align}
With the relations between the couplings given in~\cite{Schrempp:1996fb}, one analytically obtains
\begin{align}
g^{\prime 2} (\tau) &= \frac{c_1}{d_1 - \tau}\;, \ c_1 = \frac{80 \pi^2}{41}\;, \ d_1 = \frac{20 \pi (1-\sin^2 \theta_W (\tau_Z))}{41 \alpha_{em} (\tau_Z)} + \tau_Z \;, \\
g^2 (\tau) &= \frac{c_2}{d_2 + \tau}\;, \ c_2 = \frac{48 \pi^2}{19}\;, \ d_2 = \frac{12 \pi \sin^2 \theta_W (\tau_Z)}{19 \alpha_{em} (\tau_Z)} - \tau_Z \;, \\
g_s^2 (\tau) &= \frac{c_3}{d_3 + \tau}\;, \ c_3 = \frac{8 \pi^2}{7}\;, \ d_3 = \frac{2 \pi}{7 \alpha_{s} (\tau_Z)} - \tau_Z \;,
\end{align}
which we then use to solve the RGEs for $h_t$ and $\lambda_\phi$ numerically.

\bibliographystyle{JHEP}
\bibliography{lepto_sm.bib}

\providecommand{\href}[2]{#2}\begingroup\raggedright\begin{thebibliography}{10}

\bibitem{Kuzmin:1985mm}
V.~A. Kuzmin, V.~A. Rubakov and M.~E. Shaposhnikov, \emph{{On the Anomalous
  Electroweak Baryon Number Nonconservation in the Early Universe}},
  \href{https://doi.org/10.1016/0370-2693(85)91028-7}{\emph{Phys. Lett.}
  {\bfseries 155B} (1985) 36}.

\bibitem{Huet:1994jb}
P.~Huet and E.~Sather, \emph{{Electroweak baryogenesis and standard model CP
  violation}}, \href{https://doi.org/10.1103/PhysRevD.51.379}{\emph{Phys. Rev.}
  {\bfseries D51} (1995) 379}
  [\href{https://arxiv.org/abs/hep-ph/9404302}{{\ttfamily hep-ph/9404302}}].

\bibitem{Patrignani:2016xqp}
{\scshape Particle Data Group} collaboration, \emph{{Review of Particle
  Physics}}, \href{https://doi.org/10.1088/1674-1137/40/10/100001}{\emph{Chin.
  Phys.} {\bfseries C40} (2016) 100001}.

\bibitem{Buchmuller:1994qy}
W.~Buchm{\"u}ller and O.~Philipsen, \emph{{Phase structure and phase transition
  of the SU(2) Higgs model in three-dimensions}},
  \href{https://doi.org/10.1016/0550-3213(95)00124-B}{\emph{Nucl. Phys.}
  {\bfseries B443} (1995) 47}
  [\href{https://arxiv.org/abs/hep-ph/9411334}{{\ttfamily hep-ph/9411334}}].

\bibitem{Kajantie:1995kf}
K.~Kajantie, M.~Laine, K.~Rummukainen and M.~E. Shaposhnikov, \emph{{The
  Electroweak phase transition: A Nonperturbative analysis}},
  \href{https://doi.org/10.1016/0550-3213(96)00052-1}{\emph{Nucl. Phys.}
  {\bfseries B466} (1996) 189}
  [\href{https://arxiv.org/abs/hep-lat/9510020}{{\ttfamily hep-lat/9510020}}].

\bibitem{Csikor:1998ew}
F.~Csikor, Z.~Fodor and J.~Heitger, \emph{{Where does the hot electroweak phase
  transition end?}},
  \href{https://doi.org/10.1016/S0920-5632(99)85166-4}{\emph{Nucl. Phys. Proc.
  Suppl.} {\bfseries 73} (1999) 659}
  [\href{https://arxiv.org/abs/hep-ph/9809293}{{\ttfamily hep-ph/9809293}}].

\bibitem{Fukugita:1986hr}
M.~Fukugita and T.~Yanagida, \emph{{Baryogenesis Without Grand Unification}},
  \href{https://doi.org/10.1016/0370-2693(86)91126-3}{\emph{Phys. Lett.}
  {\bfseries B174} (1986) 45}.

\bibitem{Blanchet:2012bk}
S.~Blanchet and P.~Di~Bari, \emph{{The minimal scenario of leptogenesis}},
  \href{https://doi.org/10.1088/1367-2630/14/12/125012}{\emph{New J. Phys.}
  {\bfseries 14} (2012) 125012}
  [\href{https://arxiv.org/abs/1211.0512}{{\ttfamily 1211.0512}}].

\bibitem{tHooft:1976rip}
G.~'t~Hooft, \emph{{Symmetry Breaking Through Bell-Jackiw Anomalies}},
  \href{https://doi.org/10.1103/PhysRevLett.37.8}{\emph{Phys. Rev. Lett.}
  {\bfseries 37} (1976) 8}.

\bibitem{tHooft:1976snw}
G.~'t~Hooft, \emph{{Computation of the Quantum Effects Due to a
  Four-Dimensional Pseudoparticle}},
  \href{https://doi.org/10.1103/PhysRevD.18.2199.3,
  10.1103/PhysRevD.14.3432}{\emph{Phys. Rev.} {\bfseries D14} (1976) 3432}.

\bibitem{PhysRevLett.44.912}
R.~N. Mohapatra and G.~Senjanovi\ifmmode~\acute{c}\else \'{c}\fi{},
  \emph{Neutrino mass and spontaneous parity nonconservation},
  \href{https://doi.org/10.1103/PhysRevLett.44.912}{\emph{Phys. Rev. Lett.}
  {\bfseries 44} (1980) 912}.

\bibitem{Biondini:2017rpb}
S.~Biondini et~al., \emph{{Status of rates and rate equations for thermal
  leptogenesis}}, \href{https://doi.org/10.1142/S0217751X18420046}{\emph{Int.
  J. Mod. Phys. A} {\bfseries 33} (2018) 1842004}
  [\href{https://arxiv.org/abs/1711.02864}{{\ttfamily 1711.02864}}].

\bibitem{Drewes:2017zyw}
M.~Drewes, B.~Garbrecht, P.~Hernandez, M.~Kekic, J.~Lopez-Pavon, J.~Racker
  et~al., \emph{{ARS Leptogenesis}},
  \href{https://doi.org/10.1142/S0217751X18420022}{\emph{Int.\ J.\ Mod.\ Phys.\
  A} {\bfseries 33} (2018) 1842002}
  [\href{https://arxiv.org/abs/1711.02862}{{\ttfamily 1711.02862}}].

\bibitem{Chun:2017spz}
E.~Chun et~al., \emph{{Probing Leptogenesis}},
  \href{https://doi.org/10.1142/S0217751X18420058}{\emph{Int.\ J.\ Mod.\ Phys.\
  A} {\bfseries 33} (2018) 1842005}
  [\href{https://arxiv.org/abs/1711.02865}{{\ttfamily 1711.02865}}].

\bibitem{Hagedorn:2017wjy}
C.~Hagedorn, R.~Mohapatra, E.~Molinaro, C.~Nishi and S.~Petcov, \emph{{CP
  Violation in the Lepton Sector and Implications for Leptogenesis}},
  \href{https://doi.org/10.1142/S0217751X1842006X}{\emph{Int.\ J.\ Mod.\ Phys.\
  A} {\bfseries 33} (2018) 1842006}
  [\href{https://arxiv.org/abs/1711.02866}{{\ttfamily 1711.02866}}].

\bibitem{Dev:2017trv}
P.~S.~B. Dev, P.~Di~Bari, B.~Garbrecht, S.~Lavignac, P.~Millington and
  D.~Teresi, \emph{{Flavor effects in leptogenesis}},
  \href{https://doi.org/10.1142/S0217751X18420010}{\emph{Int.\ J.\ Mod.\ Phys.\
  A} {\bfseries 33} (2018) 1842001}
  [\href{https://arxiv.org/abs/1711.02861}{{\ttfamily 1711.02861}}].

\bibitem{Dev:2017wwc}
B.~Dev, M.~Garny, J.~Klaric, P.~Millington and D.~Teresi, \emph{{Resonant
  enhancement in leptogenesis}},
  \href{https://doi.org/10.1142/S0217751X18420034}{\emph{Int.\ J.\ Mod.\ Phys.\
  A} {\bfseries 33} (2018) 1842003}
  [\href{https://arxiv.org/abs/1711.02863}{{\ttfamily 1711.02863}}].

\bibitem{Anisimov:2010dk}
A.~Anisimov, W.~Buchmüller, M.~Drewes and S.~Mendizabal, \emph{{Quantum
  Leptogenesis I}}, \href{https://doi.org/10.1016/j.aop.2011.02.002,
  10.1016/j.aop.2013.05.00}{\emph{Annals Phys.} {\bfseries 326} (2011) 1998}
  [\href{https://arxiv.org/abs/1012.5821}{{\ttfamily 1012.5821}}].

\bibitem{Anisimov:2010aq}
A.~Anisimov, W.~Buchm{\"u}ller, M.~Drewes and S.~Mendizabal,
  \emph{{Leptogenesis from Quantum Interference in a Thermal Bath}},
  \href{https://doi.org/10.1103/PhysRevLett.104.121102}{\emph{Phys. Rev. Lett.}
  {\bfseries 104} (2010) 121102}
  [\href{https://arxiv.org/abs/1001.3856}{{\ttfamily 1001.3856}}].

\bibitem{Anisimov:2010gy}
A.~Anisimov, D.~Besak and D.~B{\"o}deker, \emph{{Thermal production of
  relativistic Majorana neutrinos: Strong enhancement by multiple soft
  scattering}},
  \href{https://doi.org/10.1088/1475-7516/2011/03/042}{\emph{JCAP} {\bfseries
  1103} (2011) 042} [\href{https://arxiv.org/abs/1012.3784}{{\ttfamily
  1012.3784}}].

\bibitem{Bodeker:2013qaa}
D.~B{\"o}deker and M.~W{\"o}rmann, \emph{{Non-relativistic leptogenesis}},
  \href{https://doi.org/10.1088/1475-7516/2014/02/016}{\emph{JCAP} {\bfseries
  02} (2014) 016} [\href{https://arxiv.org/abs/1311.2593}{{\ttfamily
  1311.2593}}].

\bibitem{Bodeker:2014hqa}
D.~B{\"o}deker and M.~Laine, \emph{{Kubo relations and radiative corrections
  for lepton number washout}},
  \href{https://doi.org/10.1088/1475-7516/2014/05/041}{\emph{JCAP} {\bfseries
  05} (2014) 041} [\href{https://arxiv.org/abs/1403.2755}{{\ttfamily
  1403.2755}}].

\bibitem{Bodeker:2017deo}
D.~B{\"o}deker and M.~Sangel, \emph{{Lepton asymmetry rate from quantum field
  theory: NLO in the hierarchical limit}},
  \href{https://doi.org/10.1088/1475-7516/2017/06/052}{\emph{JCAP} {\bfseries
  06} (2017) 052} [\href{https://arxiv.org/abs/1702.02155}{{\ttfamily
  1702.02155}}].

\bibitem{Biondini:2016arl}
S.~Biondini, N.~Brambilla and A.~Vairo, \emph{{CP asymmetry in heavy Majorana
  neutrino decays at finite temperature: the hierarchical case}},
  \href{https://doi.org/10.1007/JHEP09(2016)126}{\emph{JHEP} {\bfseries 09}
  (2016) 126} [\href{https://arxiv.org/abs/1608.01979}{{\ttfamily
  1608.01979}}].

\bibitem{Buchmuller:2002rq}
W.~Buchm{\"u}ller, P.~Di~Bari and M.~Pl{\"u}macher, \emph{{Cosmic microwave
  background, matter - antimatter asymmetry and neutrino masses}},
  \href{https://doi.org/10.1016/S0550-3213(02)00737-X,
  10.1016/j.nuclphysb.2007.11.030}{\emph{Nucl. Phys.} {\bfseries B643} (2002)
  367} [\href{https://arxiv.org/abs/hep-ph/0205349}{{\ttfamily
  hep-ph/0205349}}].

\bibitem{Buchmuller:2001sr}
W.~Buchm{\"u}ller and M.~Pl{\"u}macher, \emph{{Spectator processes and
  baryogenesis}},
  \href{https://doi.org/10.1016/S0370-2693(01)00614-1}{\emph{Phys. Lett. B}
  {\bfseries 511} (2001) 74}
  [\href{https://arxiv.org/abs/hep-ph/0104189}{{\ttfamily hep-ph/0104189}}].

\bibitem{Nardi:2005hs}
E.~Nardi, Y.~Nir, J.~Racker and E.~Roulet, \emph{{On Higgs and sphaleron
  effects during the leptogenesis era}},
  \href{https://doi.org/10.1088/1126-6708/2006/01/068}{\emph{JHEP} {\bfseries
  01} (2006) 068} [\href{https://arxiv.org/abs/hep-ph/0512052}{{\ttfamily
  hep-ph/0512052}}].

\bibitem{Garbrecht:2014kda}
B.~Garbrecht and P.~Schwaller, \emph{{Spectator Effects during Leptogenesis in
  the Strong Washout Regime}},
  \href{https://doi.org/10.1088/1475-7516/2014/10/012}{\emph{JCAP} {\bfseries
  10} (2014) 012} [\href{https://arxiv.org/abs/1404.2915}{{\ttfamily
  1404.2915}}].

\bibitem{Bellac:2011kqa}
M.~L. Bellac, \emph{{Thermal Field Theory}}. Cambridge University Press, 2011.

\bibitem{Covi:1996wh}
L.~Covi, E.~Roulet and F.~Vissani, \emph{{CP violating decays in leptogenesis
  scenarios}}, \href{https://doi.org/10.1016/0370-2693(96)00817-9}{\emph{Phys.
  Lett.} {\bfseries B384} (1996) 169}
  [\href{https://arxiv.org/abs/hep-ph/9605319}{{\ttfamily hep-ph/9605319}}].

\bibitem{Bergerhoff:1994sj}
B.~Bergerhoff and C.~Wetterich, \emph{{The Strongly interacting electroweak
  phase transition}},
  \href{https://doi.org/10.1016/0550-3213(95)00079-8}{\emph{Nucl. Phys.}
  {\bfseries B440} (1995) 171}
  [\href{https://arxiv.org/abs/hep-ph/9409295}{{\ttfamily hep-ph/9409295}}].

\bibitem{Philipsen:1996af}
O.~Philipsen, M.~Teper and H.~Wittig, \emph{{On the mass spectrum of the SU(2)
  Higgs model in (2+1)- dimensions}},
  \href{https://doi.org/10.1016/0550-3213(96)00156-3}{\emph{Nucl. Phys.}
  {\bfseries B469} (1996) 445}
  [\href{https://arxiv.org/abs/hep-lat/9602006}{{\ttfamily hep-lat/9602006}}].

\bibitem{Philipsen:1997rq}
O.~Philipsen, M.~Teper and H.~Wittig, \emph{{Scalar gauge dynamics in
  (2+1)-dimensions at small and large scalar couplings}},
  \href{https://doi.org/10.1016/S0550-3213(98)00330-7}{\emph{Nucl. Phys.}
  {\bfseries B528} (1998) 379}
  [\href{https://arxiv.org/abs/hep-lat/9709145}{{\ttfamily hep-lat/9709145}}].

\bibitem{Kapusta:2006pm}
J.~I. Kapusta and C.~Gale, \emph{{Finite-temperature field theory: Principles
  and applications}}, Cambridge Monographs on Mathematical Physics. Cambridge
  University Press, 2011,
  \href{https://doi.org/10.1017/CBO9780511535130}{10.1017/CBO9780511535130}.

\bibitem{Besak:2010fb}
D.~Besak and D.~B{\"o}deker, \emph{{Hard Thermal Loops for Soft or Collinear
  External Momenta}},
  \href{https://doi.org/10.1007/JHEP05(2010)007}{\emph{JHEP} {\bfseries 05}
  (2010) 007} [\href{https://arxiv.org/abs/1002.0022}{{\ttfamily 1002.0022}}].

\bibitem{Arnold:2002ja}
P.~B. Arnold, G.~D. Moore and L.~G. Yaffe, \emph{{Photon and gluon emission in
  relativistic plasmas}},
  \href{https://doi.org/10.1088/1126-6708/2002/06/030}{\emph{JHEP} {\bfseries
  06} (2002) 030} [\href{https://arxiv.org/abs/hep-ph/0204343}{{\ttfamily
  hep-ph/0204343}}].

\bibitem{Carrington:1991hz}
M.~E. Carrington, \emph{{The Effective potential at finite temperature in the
  Standard Model}}, \href{https://doi.org/10.1103/PhysRevD.45.2933}{\emph{Phys.
  Rev.} {\bfseries D45} (1992) 2933}.

\bibitem{contributors-gsl-gnu-2010}
G.~P. Contributors, ``{GSL} - {GNU} scientific library - {GNU} project - free
  software foundation {(FSF)}.'' http://www.gnu.org/software/gsl/, 2010.

\bibitem{schling2014boost}
B.~Schäling, \emph{The boost C++ libraries}. XML Press, 2014.

\bibitem{Schrempp:1996fb}
B.~Schrempp and M.~Wimmer, \emph{{Top quark and Higgs boson masses: Interplay
  between infrared and ultraviolet physics}},
  \href{https://doi.org/10.1016/0146-6410(96)00059-2}{\emph{Prog. Part. Nucl.
  Phys.} {\bfseries 37} (1996) 1}
  [\href{https://arxiv.org/abs/hep-ph/9606386}{{\ttfamily hep-ph/9606386}}].

\bibitem{Arason:1991ic}
H.~Arason, D.~J. Castano, B.~Keszthelyi, S.~Mikaelian, E.~J. Piard, P.~Ramond
  et~al., \emph{{Renormalization group study of the standard model and its
  extensions. 1. The Standard model}},
  \href{https://doi.org/10.1103/PhysRevD.46.3945}{\emph{Phys. Rev.} {\bfseries
  D46} (1992) 3945}.

\end{thebibliography}\endgroup

\end{document}